\documentclass[journal=jctcce,manuscript=article]{achemso}
\usepackage[version=3]{mhchem} 

\usepackage{physics} 
\usepackage{bm}
\usepackage[nameinlink, capitalize]{cleveref}
\usepackage{amssymb}
\usepackage{booktabs}
\usepackage{braket}

\newcommand*{\imag}{\text{i}}
\newcommand*{\e}[1]{\text{e}^{#1}}
\newcommand*{\au}{\,\mathrm{a.u.}}
\newcommand*{\llangle}{\langle\!\langle}
\newcommand*{\rrangle}{\rangle\!\rangle}
\DeclareMathOperator*{\argmin}{argmin}

\author{Eirill Hauge}
\affiliation[UiO]{Hylleraas Centre for Quantum Molecular Sciences, Department of Chemistry, University of Oslo, P.O. Box 1033, Blindern, 0315 Oslo, Norway}
\alsoaffiliation[Simula]{Department of Numerical Analysis and Scientific Computing, Simula Research Laboratory,  Kristian Augusts gate 23, 0164 Oslo, Norway}

\author{H{\aa}kon Emil Kristiansen}
\affiliation[UiO]{Hylleraas Centre for Quantum Molecular Sciences, Department of Chemistry, University of Oslo, P.O. Box 1033, Blindern, 0315 Oslo, Norway}

\author{Lukas Konecny}
\affiliation[UiT]
{Hylleraas Centre for Quantum Molecular Sciences, Department of Chemistry, University of Troms\o---The Arctic University of Norway, N-9037 Troms\o, Norway}
\alsoaffiliation[Max Planck]{Max Planck Institute for the Structure and Dynamics of Matter, Center for Free Electron Laser Science, Luruper Chaussee 149, 22761 Hamburg, Germany}

\author{Marius Kadek}
\affiliation[UiT]
{Hylleraas Centre for Quantum Molecular Sciences, Department of Chemistry, University of Troms\o---The Arctic University of Norway, N-9037 Troms\o, Norway}
\alsoaffiliation[Northeastern]{Department of Physics, Northeastern University, Boston, Massachusetts 02115, USA}

\author{Michal Repisky}
\affiliation[UiT]
{Hylleraas Centre for Quantum Molecular Sciences, Department of Chemistry, University of Troms\o---The Arctic University of Norway, N-9037 Troms\o, Norway}
\alsoaffiliation[CU]{Department of Physical and Theoretical Chemistry, Faculty of Natural Sciences, Comenius University, Ilkovicova 6, SK-84215 Bratislava, Slovakia}

\author{Thomas Bondo Pedersen}
\affiliation[UiO]{Hylleraas Centre for Quantum Molecular Sciences, Department of Chemistry, University of Oslo, P.O. Box 1033, Blindern, 0315 Oslo, Norway}
\email{t.b.pedersen@kjemi.uio.no}

\title[dipole extrapolation]{Cost-Efficient High-Resolution Linear Absorption Spectra Through Extrapolating the Dipole Moment from Real-Time Time-Dependent Electronic-Structure Theory}


\begin{document}

\begin{abstract}
We present a novel function fitting method for approximating the propagation of the time-dependent electric dipole moment from real-time electronic structure calculations. Real-time calculations of the electronic absorption spectrum require discrete Fourier transforms of the electric dipole moment. The spectral resolution is determined by the total propagation time, i.e. the trajectory length of the dipole moment, causing a high computational cost. Our developed method uses function fitting on shorter trajectories of the dipole moment, achieving arbitrary spectral resolution through extrapolation. Numerical testing shows that the fitting method can reproduce high-resolution spectra using short dipole trajectories. The method converges with as little as $100\au$ dipole trajectories for some systems, though the difficulty converging increases with the spectral density. We also introduce an error estimate of the fit, reliably assessing its convergence and hence the quality of the approximated spectrum.  
\end{abstract}

\section{Introduction}

The rapid advancement of laser technology in the past decades allows us to probe matter on spatiotemporal scales that approach the
characteristic time and length scales of the electron, opening the field of attosecond science~\cite{Corkum2007,Nisoli2017}.
This development has forced quantum chemists to shift their attention
from the time-independent to the time-dependent Schr{\"o}dinger and Dirac
equations~\cite{li_real-time_2020,lode_colloquium_2020,ofstad_time-dependent_2023}.
Numerical approaches to laser-driven electron dynamics are often labelled \emph{real-time} methods to distinguish
them from the response-theoretical methods to the time-dependent Schr{\"o}dinger/Dirac equation, the latter solving the equations of motion perturbatively in the frequency domain~\cite{olsen_linear_1985,Helgaker2012}.

Even without explicit reference to results derived from perturbation theory such as, e.g., Fermi's golden rule, it is still possible to
extract linear and low-order nonlinear optical properties from nonperturbative real-time simulations,
including electronic absorption spectra and time-resolved pump-probe absorption spectra
that would be hard or impossible to compute using response 
theory---see Refs.~\citenum{repisky_excitation_2015,Nascimento2019,Skeidsvoll2020,Guandalini2021,moitra_accurate_2023}
for recent examples.

In this work, we focus on electronic absorption spectra extracted from electron-dynamics simulations driven by a Dirac-delta impulse,
which excites the molecule into all dipole-allowed excited states simultaneously~\cite{repisky_excitation_2015}.
Due to the nonperturbative nature of real-time methods,
the resulting spectrum contains nonlinear (e.g., two-photon) as well as linear absorption lines~\cite{Guandalini2021}. For weak pulses, the nonlinear effects are small, and the absorption spectrum is dominated by linear lines.
In practice, the induced electric dipole moment is recorded in the
course of the simulation and subsequently transformed to the frequency domain to yield the absorption cross section. When using the
conventional discrete Fourier transform to process the signal, the spectral resolution
is inversely proportional to the number of time steps $N$ and the time-step length $\Delta t$, as $\Delta\omega = 2\pi /(N \Delta t)$.
Obtaining sufficient spectral resolution typically requires tens to hundreds of thousands of time steps since $\Delta t$
cannot be increased beyond a certain limit if rapid oscillations of the electron density are to be captured.
Moreover, increasing $\Delta t$ reduces the accuracy and stability of the numerical integration scheme used to propagate
the electronic degrees of freedom.
As the computational effort in each time step requires multiple rebuilds of the Hamiltonian matrix, it is comparable to several iterations of a ground-state optimization within the chosen electronic-structure model~\cite{comment}. Hence, there is considerable interest in decreasing the number of time steps required to achieve sufficient spectral resolution.

In addition to reducing the number of time steps, it is possible to increase the computational efficiency of real-time electronic structure methods by disregarding negligible basis functions~\cite{Han2023}, 
basis-function pairs and quartets~\cite{Black2023}. As a result, in a large molecule, although there are $\mathcal{O}(N^{4})$ electron repulsion integrals (ERIs) in total, it can be shown that only $\mathcal{O}(N^{2})$ of them are significant, where $N$ refers to the number of basis functions. As shown within 
real-time time-dependent density-functional theory (RT-TDDFT),
a large prefactor associated with the evaluation of non-negligible ERIs can further be reduced by using, e.g., the resolution-of-the-identity method~\cite{Konecny2018}. By applying a spatial truncation radius upon the time-dependent density matrix, RT-TDDFT can approach the linear $\mathcal{O}(N)$ scaling~\cite{Yokojima1998,ORourke2015}.

Previous efforts to improve the spectral resolution have been made by estimating excitation energies through various signal processing
techniques~\cite{mandelshtam_harmonic_1997, roy_novel_1991, wall_extraction_1995}. More recently,
\citeauthor{bruner_accelerated_2016}~\cite{bruner_accelerated_2016}
investigated the use of Padé approximants to interpolate the discrete Fourier transforms used for the absorption spectrum.
These are all methods operating in the frequency domain, leaving no other validation options than comparison with a fully propagated spectrum. 

The original periodic signal is typically damped using a decaying exponential function to reduce unwanted artefacts arising when the discrete Fourier transform is applied to oscillating functions in simulations with finite trajectory length. In the time domain, the number of time points can be increased by padding the damped signal with zeros, leading to finer spectra. However, this artificial extension of the trajectory length can only be applied on sufficiently damped signals.

In this work, we investigate a
more sophisticated and powerful alternative: the extrapolation of a short signal.
The discrete Fourier transform of an extrapolated signal achieves increasingly higher spectral resolution as the extrapolation length increases.
This requires the development of a stable and reliable method for time-series forecasting.
The inherently harmonic character of the time-dependent wave function in the absence of an external field suggests that such forecasting of molecular properties should be possible.
Importantly, the forecasted signal can be verified in the time domain by comparing it with relatively few additional time steps.
To the best of our knowledge, no published work exists on improving the spectral resolution by such extrapolation of
the time-dependent dipole moment. 

The current success and popularity of machine learning is undeniable, including use cases in chemistry~\cite{dral_quantum_2020,hase_how_2019,butler_machine_2018,kitchin_machine_2018,Chen2018,Secor2021, Ullah2022}, and one might be tempted to leverage artificial neural networks for forecasting the time-dependent electric-dipole
moment. 
However, while artificial neural networks are powerful tools for pattern detection in large data sets, they struggle
with precise and reliable extrapolations~\cite{xu_how_2021,oleinik_what_2019}. 
Although the universal approximation theorem~\cite{pinkus_approximation_1999} tells us that an excellent \emph{interpolation} can be achieved,
it does not guarantee a stable \emph{extrapolation}.
In order to achieve a stable extrapolation, over-fitting must be avoided by enforcing sufficient restrictions.

In this paper, we present a novel approach for obtaining high-resolution absorption spectra from real-time simulations of laser-driven electron dynamics
by exploiting \emph{a priori} knowledge of the form of the dipole function from quantum mechanics in a finite-dimensional Hilbert space.
The form of the dipole function thus is motivated by the underlying physics, with unknown parameters to be determined by fitting
a short dipole trajectory from a real-time simulation.
The fitted function may be evaluated at any point in time, meaning that it can be extrapolated in the
time domain to arbitrary future time. This further implies that we can achieve arbitrary spectral resolution. For sufficiently weak Dirac-delta impulse, the evaluation of absorption spectra based on these fitted functions
may use analytical expressions for the linear response function~\cite{olsen_linear_1985}.

Working in the time domain, a quantitative error measure of the fitted dipole function can be monitored during the course of the real-time simulation
and used to evaluate convergence.
This way, an unnecessarily long real-time propagation can be avoided by automatically terminating the propagation upon the convergence of the fit.
The developed method is independent of the quantum mechanical model and is tested with several molecular systems using mainly RT-TDDFT~\cite{Runge1984,VanLeeuwen1999,theilhaber_ab_1992,yabana_time-dependent_1996,Lopata2011,repisky_excitation_2015,Wibowo2021}.
Despite certain flaws arising mainly from the almost universally adopted adiabatic density-functional
approximation~\cite{li_real-time_2020,provorse_electron_2016}, RT-TDDFT
is the far most widely used electronic-structure method for laser-driven electron dynamics.
With computational costs comparable to (or \emph{below}) time-dependent Hartree-Fock theory~\cite{Li2005},
RT-TDDFT takes into account electron-correlation effects
that would otherwise require advanced and computationally expensive wave-function theories~\cite{lode_colloquium_2020,ofstad_time-dependent_2023}.
To demonstrate the independence of the underlying electronic-structure theory, we also present results obtained from real-time
time-dependent configuration interaction singles (RT-TDCIS)~\cite{Foresman1992,Rohringer2006,Greenman2010} theory.

We will start with a short presentation of the electric-dipole approximation within real-time simulations
before introducing the proposed method for fitting the time-dependent electric-dipole moment. 
After briefly laying out the simulation details for the real-time simulation of a selection of systems,
the results of the fitting method on these systems are presented and discussed.
Finally, we reflect on the performance of the fitting method and discuss potential future improvements.

\section{Theory} 
In this work, we employ the following conventions:
Subscripts $u$, $v$ denote Cartesian components,
vectors are typed in boldface, and quantum-mechanical operators are denoted by a hat.
Following the convention of response theory by \citeauthor{olsen_linear_1985}~\cite{olsen_linear_1985},
we define the Fourier transform and its inverse according to
\begin{align}\label{eq:Fourier}
    \tilde{f}(\omega) 
    &= \mathcal{F}[f(t)]
    = \frac{1}{2\pi}\int^{\infty}_{-\infty}
     f(t) 
     \e{\imag\omega t} \dd t
     , \\
   f(t)
    &= \mathcal{F}^{-1}[\tilde{f}(\omega)]
    = \int^{\infty}_{-\infty}
     \tilde{f}(\omega) 
     \e{-\imag\omega t} \dd \omega,
\end{align} 
where the transformed function is denoted by a tilde.
Atomic units are used throughout unless otherwise specified.

\subsection{Real-time simulations of absorption spectra}

Within the clamped-nucleus Born-Oppenheimer approximation, real-time simulations of electronic absorption spectra typically assume the electric-dipole approximation, where a molecule is subjected to a time-dependent spatially uniform electric field, $F(t)$.
The time-dependent Hamiltonian reads
\begin{equation}\label{eq:H}
    \hat{H}(t) = \hat{H}_0 + \hat{V}(t),
\end{equation}
where $\hat{H}_0$ is the time-independent electronic Hamiltonian, and the interaction operator is given by
\begin{equation}\label{eq:V}
    \hat{V}(t) = - \bm{\hat\mu} \cdot \vb{u} F(t),
\end{equation}
where $\bm{\hat\mu}$ is the electric dipole moment operator.
The linear polarization direction of the electric field is determined by the real unit vector $\vb{u}$, such that the field aligns with one of the Cartesian axes.  
This implies the form $\hat{V}(t) = -\hat{\mu}_u F(t)$, where $\hat{\mu}_u$ is the component of $\bm{\hat\mu}$ along the polarization direction.

We assume that the electronic system is in the ground state $\ket{0}$ at time $t < 0$, and that the external field $F(t)$ is only active between $t = 0$ and time $t_0 \geq 0$. At time $t_0$, the Hamiltonian reduces to the time-independent Hamiltonian such that Schrödinger's equation for $t \geq t_0$ becomes 
\begin{equation}\label{eq:Schro}
    \hat{H}_0 \ket{\Psi(t)}
    = \imag \frac{\mathrm{d}}{\mathrm{d} t} \ket{\Psi(t)}.
\end{equation}
The time-dependent wave function in the absence of the external field oscillates around the solution at time $t_0$, $\ket{\Psi(t_0)} = \sum_n k_n(t_0) \ket{n}$, as given by
\begin{equation}\label{eq:Psi_exact}
    \ket{\Psi(t)} = \e{-\imag \hat{H}_0(t - t_0)} \ket{\Psi(t_0)} 
    = \sum_n k_n(t_0) \e{-\imag E_n (t - t_0)}  \ket{n},
\end{equation}
where $\ket{n}$ denotes a normalized eigenfunction of the unperturbed Hamiltonian,
$\hat{H}_0 \ket{n} = E_n \ket{n}$~\cite{helgaker_book_2000, GriffithsDavidJ2017Itqm}.
This formulation is exact when the electronic continuum is excluded, e.g., by choosing a fixed, finite basis as commonly done in quantum chemistry.
Actual simulations are not performed in the energy eigenbasis but in, e.g., a basis of Slater determinants, implying that the
coefficients $k_n(t_0)$ are not known.

In order to obtain the electronic absorption spectrum averaged over all molecular orientations relative to the
electric field, the time-dependent electric dipole moment
$\mu_u(t) = \braket{\Psi(t) \vert \hat{\mu}_u \vert \Psi(t)}$
is calculated in three separate simulations
with the electric field polarized in each of the three Cartesian directions ($u = x, y, z$).
The absorption cross-section is then obtained from the Fourier transform of the dipole moments, $\tilde{\mu}_u(\omega)$, as~\cite{Tannor2007}
\begin{equation}\label{eq:S}
    S(\omega) = \frac{4 \pi \omega}{3 c}\Im\left[
      \frac{\tilde\mu_x(\omega)\tilde{F}^*(\omega)}{|\tilde{F}(\omega)|^2}
    + \frac{\tilde\mu_y(\omega)\tilde{F}^*(\omega)}{|\tilde{F}(\omega)|^2}
    + \frac{\tilde\mu_z(\omega)\tilde{F}^*(\omega)}{|\tilde{F}(\omega)|^2}
    \right],
\end{equation}
where $c$ is the speed of light. The resulting spectrum contains both linear (one-photon
transitions between the ground and excited states) and nonlinear (multi-photon transitions between the ground and excited states,
and one- and multi-photon transitions between excited states) absorptions, as recently stressed by \citeauthor{Guandalini2021}~\cite{Guandalini2021}
We note that only the \emph{induced} dipole moment, that is the total dipole moment with the static ground-state part subtracted, contributes to the absorption cross section but, for notational convenience,
we will only distinguish between that and the total dipole moment when it is strictly required.

Since the dipole moment is calculated on a finite discrete time grid, the Fourier transforms are replaced by
discrete Fourier transforms, thus introducing artificial periodic boundary conditions.
To avoid artefacts from these,
the dipole moment is multiplied by a damping factor before the discrete Fourier transform, i.e.,
\begin{equation}
    \tilde\mu_{u}(\omega) = \mathcal{F}[\mu_u(t)\e{-\gamma\vert t\vert}]
    = \frac{1}{2\pi} \int_{0}^\infty \mu(t) \e{\imag(\omega + \imag \gamma)t} \dd t,
\end{equation}
where we have used that the induced dipole moment vanishes for $t<0$. The Fourier transform thus becomes a Laplace transform.
The parameter $\gamma \in \mathbb{R}_+$ 
can be interpreted as a common (inverse) lifetime of all excited states, giving rise to Lorentzian line shapes
in the simulated absorption spectra~\cite{norman_near-resonant_2001}. 
The discrete Fourier transform, however, requires a very large, often prohibitive, number of
time steps to achieve sufficient spectral resolution. In the following sections, we will describe an extrapolation technique
aiming at high resolution with short simulation time.

\subsection{The expected form of the electric dipole moment}

Once the external field is turned off, the time-dependent electric dipole moment 
evolves according to $\mu_u(t) = \bra{\Psi(t)}\hat\mu_u\ket{\Psi(t)}$, where $\ket{\Psi(t)}$ is defined in \cref{eq:Psi_exact}. The dipole moment 
oscillates with the Bohr frequencies $\omega_{n m} = E_{n} - E_{m}$ according to
\begin{align}\label{eq:full_dipole_moment}
\begin{split}
    \mu_u(t)
    =& 2 \sum_{n > m} \Bigg\{ 
        \Re\left[\bra{n}\hat\mu_u\ket{m} k_n^*(t_0)k_m(t_0)\right]\cos(\omega_{n m}(t - t_0)) \\
        &- \Im\left[\bra{n}\hat\mu_u\ket{m} k_n^*(t_0)k_m(t_0)\right] \sin(\omega_{n m}(t - t_0)) 
        \Bigg\} \\
        &+ \sum_{n} |k_n(t_0)|^2 \bra{n}\hat\mu_u\ket{n},
\end{split}
\end{align}
for time $t \geq t_0$~\cite{hauge_extrapolating_2021}. 
The function form of the approximated dipole moment $ \bar\mu_u(t) \approx \mu_u(t)$ will therefore be given by
\begin{equation}\label{eq:mu_tilde}
    \bar\mu_u(t) = c_0^u + \sum_{i = 1}^{N_\omega^u} \left[
         c_{i}^u\sin(\omega_i^u (t - t_0)) + c^u_{N_\omega^u + i}\cos(\omega_i^u (t - t_0))
         \right],
\end{equation}
where $N_\omega^u$ is the number of participating frequencies $\omega_i^u$, each frequency with two independent linear coefficients $c_{i}^u$ and $c^u_{N_\omega^u + i}$.
If we can determine these frequencies and their corresponding real coefficients
from a short dipole time series, we obtain a \emph{continuous} dipole function and, hence,
infinite spectral resolution.

As will be described in detail below, we estimate the participating frequencies using the poles of a Fourier-Padé approximant, while the
linear coefficients are determined using linear regression in a subsequent step.

\subsection{Estimating Bohr frequencies}
In order to estimate the frequencies of the dipole moment, we will investigate the singular points of the Fourier-Padé approximant,
originally introduced in real-time quantum simulations by \citeauthor{bruner_accelerated_2016}~\cite{bruner_accelerated_2016}
In general, the Padé approximant is used to accelerate the convergence of a truncated power series. The discrete Fourier transform can be written as the power series
\begin{equation}
    \tilde{\mu}_u(\omega) = \frac{\Delta t}{2 \pi}\sum_{n = 0}^{N_t - 1} \mu_u(t_n) z^n,
\end{equation}
where $z$ depends on the frequency according to
\begin{equation}
    z \equiv z(\omega) = \e{(\imag\omega - \gamma)\Delta t}.
\end{equation} 
The diagonal Fourier-Padé approximates the Fourier transform using two polynomials $P_u(z)$ and $Q_u(z)$
of degree $M = (N_t-1)/2$,
\begin{equation} 
    [M/M]_{\mu_u}(\omega) = \frac{\Delta t}{2 \pi}\frac{P_u(z(\omega))}{Q_u(z(\omega))},
\end{equation}
where the coefficients of the polynomials create a Toeplitz linear system. For details see Ref.~\citenum{bruner_accelerated_2016}.
The Fourier-Padé poles, denoted $z_p^u$, are found by
\begin{equation}
    Q_u(z_p^u) = 0,
\end{equation}
where the damping parameter $\gamma$ is set to zero, 
as the damping parameter removes the singularities of the spectrum.
The Bohr frequencies are positive and real-valued, while the frequencies corresponding to roots of $Q_u(z)$ will be complex. The number of roots of $Q_u(z)$, (which amounts to $M$ roots), should also significantly exceed the number of Bohr frequencies. The potential frequencies are given by
\begin{align}\label{eq:omega_p}
    \omega_p^u = \left|\frac{\ln(z_p^u)}{\Delta t} \right|,
\end{align}
where $\ln(z_p^u)$ returns the principal value of the logarithm. 
Only roots $\Im(z_p^u) > 0$ are considered, as the complex conjugate root theorem states that complex roots will form conjugate pairs. 
These conjugate pairs yield duplicates of the real-valued frequencies. 
Real-valued roots $z_p^u$ yield purely imaginary frequencies, and are therefore also excluded. The potential frequencies $\omega_p^u$  discard the imaginary component and should represent extrema of the Fourier-Padé spectrum, not singular points like $z_p^u$.

Estimating the potential frequencies uses the Python NumPy~\cite{numpy} library to compute the eigenvalues of the companion matrix~\cite{companion_matrix} of the polynomial $Q_u(z)$ to determine its roots. This method exhibits poor scaling with respect to the number of time points $N_t$, representing a computational bottleneck of the dipole-moment $\bar\mu_u(t)$ fitting procedure. In real-time simulations using very small time steps, one may safely increase the step length on the dipole data used to create the Fourier-Padé approximant to alleviate the computational cost.
As shown by \citeauthor{mattiat_efficient_2018}~\cite{mattiat_efficient_2018}, the convergence of the Fourier-Padé approximant is mostly impacted by the trajectory length $N_t\Delta t$, and not the time step itself. However, the discrete Fourier transform, and hence also the Fourier Padé, is periodic with a cycle length of $2\pi/\Delta t$. Peaks above $\pi/\Delta t$ will \emph{fold back} due to anti-symmetry and appear as negative duplicates polluting the spectrum. Therefore, it is crucial to keep the time step $\Delta t < \pi/\omega_\text{max}$, where $\omega_\text{max}$ is the largest significant frequency in the signal. 

The potential frequencies $\omega_p^u$ must be classified as either an estimated frequency or a redundant root.
The classification is based on the assumption that $\ln(z_p^u)/(\imag \Delta t)$ should have a significant
imaginary component if $\omega_p^u$ is a redundant root, while it should lie close to the real axis if $\omega_p^u$
corresponds to an actual Bohr frequency. This further means that $Q_u(z(\omega_p^u))$ should be close to zero and
that $[M/M]_{\mu_u}(\omega_p^u)$ should be large for estimated frequencies. 
Hence, we create a two-dimensional representation $\vb{r}^u_p$ of the prospective frequencies $\omega_p^u$ given by 
\begin{align}
    [r_p^u]_x &= 1 - \frac{ X(\omega_p^u)
         - \min_{\omega_q^u}\big[X(\omega_q^u)\big]}
         {\max_{\omega_q^u}\big[X(\omega_q^u)\big]
         - \min_{\omega_q^u}\big[X(\omega_q^u)\big]}, \\
    [r_p^u]_y &= \frac{Y(\omega_p^u)
         - \min_{\omega_q^u}\big[Y(\omega_q^u)\big]}
         {\max_{\omega_q^u}\big[Y(\omega_q^u)\big]
         - \min_{\omega_q^u}\big[Y(\omega_q^u)\big]} ,
\end{align}
where the unnormalized features are defined as
\begin{align}
    X(\omega_p^u) &= \log_{10}\Big(\left|[M/M]_{\mu_u}(\omega_p^u)\right|\Big), \\
    Y(\omega_p^u) &= \log_{10}\Big(\left|Q_u(z(\omega_p^u))\right|\Big)  .
\end{align}
The base-$10$ logarithm is used to manage the extreme scaling of both features, as prospective frequencies should cause $Q_u(z(\omega_p^u))$ to approach zero and hence be a nearly singular point of $[M/M]_{\mu_u}(\omega_p^u)$. The features are constructed such that estimated frequencies should be close to $\vb{r}^u_p = (0, 0)$, while redundant roots should be closer to $\vb{r}^u_p = (1, 1)$.

We use the $K$-means clustering algorithm (see e.g. \citenum{hastie_kmeans_2009,clustering}), implementation from the Python SciKit-Learn~\cite{scikit-learn} library,  with $K=2$ to classify prospective frequencies.
The $2$-means clustering algorithm is a computationally inexpensive way to separate a set into two categories. The \emph{centroid} for the cluster of potential frequencies should be closer to $(0, 0)$,
whereas the centroid for the redundancy cluster should be closer to $(1, 1)$.

\subsection{Determining the linear coefficients}

Once the frequencies are estimated, the linear coefficients are determined using linear regression. The coefficients are optimized by minimizing the \text{cost function}~\cite{hastie_linear_2009},
\begin{equation}
    R(\vb{c}^u)
    = \sum_{n = 0}^{N_t - 1} 
    \left[\mu_u(t_n) - \bar{\mu}_u(t_n; \vb{c}^u)\right]^2 .
\end{equation}
Using the general form of the dipole moment in \cref{eq:full_dipole_moment}, the linear coefficients may be optimized using a simple least squares optimization. The only restraint on the optimization of these coefficients is that they are real. This fitting procedure is general for any type of external field, $F(t)$. However,
as shown by \citeauthor{hauge_extrapolating_2021}~\cite{hauge_extrapolating_2021},
restricting the coefficients is crucial to avoid over-fitting the dipole moment.

The form $\bar\mu_u(t)$ in \cref{eq:mu_tilde} is based on the full dipole moment, correct through all orders in perturbation theory and is independent of the electric field. In this work, we will use a Dirac delta-type impulse~\cite{repisky_excitation_2015} of strength $\kappa$,
\begin{equation}
    F(t) = \kappa \delta(t),
\end{equation}
which has an infinitely wide frequency distribution and thus generates the full absorption spectrum for the given polarization direction. This implies that $t_0 = 0$. Further, we assume that the electric field strength is sufficiently weak,
such that we may regard the interaction operator $\hat{V}(t)$ as a time-dependent perturbation and assume that the interaction only induces one-photon transitions from the ground state---i.e., a linear absorption spectrum. The electric dipole moment should then be of the form $\mu_u(t) \approx \mu_u^{(0)} + \mu_u^{(1)}(t)$, 
where the zeroth order dipole moment corresponds to the ground-state value, 
$\mu_u^{(0)} = \mu_u(t = 0)$. We will now investigate an analytical expression for the first-order correction to the dipole moment induced by a weak Dirac delta impulse.

We start with the exact expression for the linear response function~\cite{olsen_linear_1985},
\begin{equation}\label{eq:linear_response}
    \llangle
        \hat{\mu}_u; \hat{V}_u(\omega) 
    \rrangle_{\omega + \imag \gamma}
    =
    - 2 \tilde{F}(\omega) \sum_{n \neq 0}\abs{\bra{0}\hat{\mu}_u\ket{n}}^2  \frac{\omega_{n 0}}
        {(\omega + \imag \gamma)^2 - \omega_{n 0}^2} ,
\end{equation}
where we have used the Fourier transform of the interaction operator $\hat{V}(t)$,
\begin{equation}
    \hat{V}_u(\omega) = \hat\mu_u \tilde{F}(\omega), \qquad
    F(\omega) = \frac{\kappa}{2\pi}.
\end{equation}
The linear response function and the first-order correction to the dipole moment are related by
$\llangle
    \hat{\mu}_u; \hat{V}_u(\omega) 
\rrangle_{\omega + \imag \gamma}
= \mathcal{F}[\mu_u^{(1)}(t)\e{-\gamma\vert t\vert}]
$. 
Since $\mu_u^{(1)}(t < 0) = 0$, we get the relation
\begin{equation}
    \int_{0}^\infty
    \mu_u^{(1)}(t) \e{-(\gamma - \imag\omega)t}\dd t
    = - 2\kappa\sum_{n \neq 0}\abs{\bra{0}\hat{\mu}_u\ket{n}}^2  \frac{\omega_{n 0}}
        {(\omega + \imag \gamma)^2 - \omega_{n 0}^2}.
\end{equation}
Using well-known Laplace transforms, it is readily verified that the first-order dipole correction must be
a linear combination of sine waves~\cite{hauge_extrapolating_2021}
\begin{equation}\label{eq:mu_ind-broad-band}
    \mu_u^{(1)}(t)= \sum_{n \neq 0}  B_n^u \sin(\omega_{n 0}t), \qquad B_n^u = 2\kappa\abs{\bra{0}\hat{\mu}_u\ket{n}}^2.
\end{equation}
We have also used that the first-order perturbation correction to the dipole moment should only include one-photon transitions. This further means that the approximated dipole moment, when using a weak Dirac delta impulse, should have the form
\begin{equation}
    \bar\mu_u(t) = c_0^u + \sum_{i = 1}^{N_\omega^u}
         c_{i}^u\sin(\omega_i^u t),
\end{equation}
where all sine coefficients are positive.

The coefficients of $\bar{\mu}_u(t)$, approximating the dipole moment from the Dirac delta impulse, are optimized using the \textit{least absolute shrinkage and selection operator (LASSO)}~\cite{Tibshirani1996} method. The coefficients are determined according to
\begin{equation}
    \vb{c}^u_\text{LASSO} =
    \argmin_{\vb{c}^u} \left\{ 
    \frac{1}{2} R(\vb{c}^u) 
    + \lambda \sum_{i} |c_i^u|
    \right\} ,
\end{equation}
where $\lambda$ is the \textit{shrinkage parameter} restricting the magnitude of the coefficients, $c_i^u$. In contrast to the
ordinary linear least-squares algorithm,
the LASSO method is iterative and therefore somewhat less computationally efficient. 
In return, this makes it possible to enforce positive coefficients, as in the implementation by SciKit-Learn~\cite{scikit-learn}. This makes the method less prone to over-fitting.

\subsection{Molecular orbital decomposition}
The electric dipole moment can be written as a sum of contributions from elementary molecular orbital (MO) transitions~\cite{repisky_excitation_2015},
\begin{equation}
    \mu_u (t) = \sum_{i a} \mu_u^{i a} (t),
\end{equation}
where $i$ and $a$ label occupied and virtual MOs, respectively.
The components $\mu_u^{i a} (t)$ are then approximated separately. 
This MO decomposition can divide a dense spectrum into a series of sparser spectra and aid in the assignment of absorption
lines~\cite{repisky_excitation_2015}.
Clustering the MO components into groups can be used to offset the increased memory consumption~\cite{ghosh_semiempirical_2019}. For the fitting method, creating clusters with well-separated frequencies could also
reduce the accumulation of errors when summing the component fits.

When fitting the individual components, the assumptions on the sign of the linear coefficients are no longer valid.
As is clear from the underlying theory and as demonstrated in practice by \citeauthor{bruner_accelerated_2016}~\cite{bruner_accelerated_2016},
the same frequencies may be found in several components $\mu_u^{i a}$, and their corresponding partial spectra may contain negative peaks.
Only the full spectrum, i.e., the sum of the components, is guaranteed to contain positive peaks exclusively.
The ordinary least squares method must therefore be used when optimizing the linear coefficients of the individual components,
which may introduce additional errors due to over-fitting in each component. 

Alternatively, the fitting algorithm may estimate the frequencies of each component separately and then optimize the linear coefficients for the full dipole moment. This way, the additional coefficient restrictions can be used in the optimization. In our experience, however,
this produces a vast number of estimated frequencies leading to problems with over-fitting even when enforcing positive linear coefficients.

\subsection{Convergence criterion}

The goal of the fitting method is to accurately construct the function $\bar{\mu}(t)$ using the shortest possible dipole trajectory.
A given trajectory is divided into two parts, a fitting domain and a verification domain. The linear coefficients are optimized using only the fitting domain, while the error is calculated on the verification domain. 
When estimating the frequencies, however, the entire available trajectory is used.
Measuring the error in the fitting domain gives the \textit{interpolation} error, which is artificially low in cases of over-fitting, whereas the error in the verification window indicates the reliability of the extrapolated dipole moment.
The error of the fit is estimated using one minus the coefficient of determination, $R^2$, i.e.,
\begin{equation}\label{eq:error}
    E_u = 1 - R^2 
    =\frac
    {\sum_n \left[\mu_u(t_n) -\bar\mu_u(t_n) \right]}
    {\sum_n \left[\mu_u(t_n) -\mu_u^m \right]},
\end{equation}
where $\mu_u^m$ is the mean value of the induced dipole moment. The error measure is unitless and independent of the magnitude of the dipole moment. The fitting method can be run in parallel with real-time simulations, which are terminated once $E_u$ drops below a pre-defined threshold value. The computational cost of the fitting method is not insignificant, and we recommend that it is run once per time intervals of $50 - 100\au$ when used to automatically terminate the real-time simulation.

Computing the error according to \cref{eq:error} provides an error estimate of the fit as a whole. In our experience from testing the algorithm with ideal multi-sinusoidal signals, the error of the fit depends primarily on the frequency estimation. Significant deviations in the estimated linear coefficients were only observed if there were frequencies missing and/or poorly estimated. In the case of ideal signals, there would only be a significant error in the fit if the frequency estimation failed. Real dipole data contains noise introduced by numerically integrating in time. How this affects the distribution of error is unknown, though it is reasonable to believe that the main source of error still lies in the frequency estimation.

The error measure $E_u$ cannot distinguish error contributions from different parts of the spectrum, preventing termination once the desired
frequency region is converged. 
In order to focus on valence excitations in the low-frequency region, we apply a low-pass (smoothing) filter to remove frequencies above a
cut-off frequency $\omega_\text{max}$ from the dipole moment in the time domain. We used a Butterworth filter, implemented by SciPy~\cite{SciPy}, which removes the high-energy part of the spectrum while leaving the lower-energy part almost unchanged. 
If bound core excitations are the main targets, a high-pass filter must be used instead.

\section{Computational details}

We test the dipole extrapolation scheme using RT-TDDFT simulations, supplemented by a few RT-TDCIS simulations to demonstrate
its applicability to wave-function-based theories.
The RT-TDDFT simulations are performed using the ReSpect program~\cite{Repisky2020}, while the
RT-TDCIS calculations are performed using the Hylleraas Quantum Dynamics (HyQD) software~\cite{HyQD}.
The RT-TDDFT and RT-TDCIS simulations are performed with analytic integration at $t = 0\au$, as described in Ref.~\citenum{repisky_excitation_2015}.
The subsequent time steps are performed numerically using the Magnus integrator for the RT-TDDFT simulations~\cite{repisky_excitation_2015}
and the three-stage Gauss-Legendre integrator~\cite{gauss_legendre_integrator} as described in Ref.~\citenum{pedersen_symplectic_2019}
with the residual norm convergence criterion $10^{-14}\au$ for the implicit equations
for the RT-TDCIS simulations.

The RT-TDCIS simulations are performed with time step $\Delta t = 0.01\au$ and field strength $\kappa = 10^{-3}\au$
The RT-TDDFT simulations for the organic molecules \ce{CH4}, \ce{CH2O}, \ce{CH3OH}, \ce{C2H6}, and \ce{C6H6}
are performed with time step $\Delta t = 0.01\au$ and field strength $\kappa = 10^{-4}\au$, while
$\Delta t = 0.01\au$ and $\kappa = 10^{-3}\au$ are used for \ce{CO2}, \ce{H2O}, and \ce{NH3}.
For the smallest systems, \ce{He}, \ce{H2}, \ce{Be}, and \ce{LiH}, $\Delta t = 0.1\au$ and $\kappa = 10^{-3}\au$ are used.

Molecular geometries are found in the supplementary information.
The simulations were performed in Dunning's cc-pVXZ and aug-cc-pVXZ, X = D,T, basis sets~\cite{Dunning1989, Kendall1992, Woon1993}
(uncontracted in the case of RT-TDDFT calculations).
The RT-TDDFT simulations were performed using the PBE0 exchange--correlation potential~\cite{slater1951, Perdew1996, Perdew1997, Adamo1999}
in the adiabatic approximation.

Simulations are performed for all three Cartesian directions for all systems, even in cases where point-group symmetry
could have been easily exploited to reduce the computational effort to one or two directions. 
While this is mainly done to make automation simple (identical treatment for all systems), 
it also provides a simple check that the extrapolation algorithm does not significantly break point-group symmetry
due to numerical noise in the input dipole trajectories.

Our implementation of the dipole-extrapolation algorithm is freely available at \url{https://github.com/HyQD/absorption-spectrum}.

\section{Results}

All reference spectra in this paper are produced from low-pass filtered electric dipole moments with a trajectory length of $4000\au$, 
such that the spectral resolution becomes $\Delta \omega = 1.6\cdot 10^{-3}\au$.
In this paper, the resolution of the fitted spectrum $\bar{S}(\omega)$ is the same as its reference spectrum. This is to allow direct comparisons of the two spectra, though the resolution of $\bar{S}(\omega)$ could be made arbitrarily fine. The Fourier transform of the approximated dipole moment $\bar{\mu}_u$ is calculated according to
\begin{equation}
        \tilde{\mu}_u(\omega) 
    =
    - \frac{1}{2\pi} \sum_{i} c_i^u \frac{\omega_{i}^u}
        {(\omega + \imag \gamma)^2 - (\omega_{i}^u)^2}.
\end{equation}
 The half-life parameter was always set to $\gamma = 0.5 \cdot 10^{-3}\pi$, and the spectra were cut at an estimated ionization energy of $0.5\au - \epsilon_\text{HOMO}$, where 
the energy of the highest occupied molecular orbital $\epsilon_\text{HOMO}$ of all systems are listed in the supplementary information. 

Using the low-pass filter will leave the lower energy part of the absorption spectrum unaltered, while the higher energy part is removed and set to zero. Differences between filtered and unfiltered spectra are shown in the supplementary information.
The low-pass filter does not give a clean cut-off at the cut-off frequency $\omega_\text{max}$ but rather a gradual lowering of the peak intensity around $\omega_\text{max}$. The cut-off frequency should therefore be set somewhat higher than the desired range of the spectrum. We have used 
$\omega_\text{max} = 4 \au$ for all systems, meaning that .

When fitting the dipole moment, the available trajectory is from when the external field is turned off at $t = 0$ to time $t = T_\text{ver}^u$.
The linear coefficients are determined on the time interval $[0, T_\text{fit}^u]$, where 
$T_\text{fit}^u = 0.75 T_\text{ver}^u$. 
The frequencies are estimated on the entire available trajectory $[0, T_\text{ver}^u]$ but with a limit on the total number of data points supplied to the Padé, set to $5\cdot 10^3$. The reduction in points, if exceeding the limit, is done by effectively increasing the time step $\Delta t$ used (by an integer factor) when creating the Padé.  
The error $E_u$ is only evaluated on $(T_\text{fit}^u, T_\text{ver}^u]$. The error in the spectrum $E_S$ is calculated the same way as in the time domain, as given in \cref{eq:error}.

\subsection{Performance on a selection of systems}

For each spatial direction, the convergence of the fit is tested every $50\au$ in time of the trajectory length, starting from $T_\text{min} = 100\au$. The simulation is terminated when the fit has converged below a given threshold or when the trajectory length reaches  $T_\text{max} = 1000\au$, which corresponds to a target minimum spectral resolution of $0.006\au$
The convergence criterion was set to $E_u < 10^{-3}$, a strict threshold corresponding to a near-perfect fit. 
The criterion was set based on preliminary investigations~\cite{hauge_extrapolating_2021}.
Since the real-time calculations on a given system using three spatial directions of the external field are independent, the trajectory length needed for a converged fit might vary between the three simulations.

The required trajectory length of each spatial direction $T_\text{ver}^u$ and their corresponding verification error $E_u$ as well as the error in the spectrum $E_S$ are listed in \cref{tab:CIS,tab:DFT}. 
The fitting of the dipole moment from RT-TDCIS calculations is shown in \cref{tab:CIS}, while the fitting of RT-TDDFT data is found in \cref{tab:DFT}. Figures of the approximated spectra of all systems can be found in the supplementary information.

\begin{table*}[ht]
    \centering
    \caption{Convergence times and corresponding errors of systems from RT-TDCIS calculations.}
    \label{tab:CIS}
    \begin{tabular}{lcccccccc}
        \toprule
         & basis & 
        $T_\text{ver}^x$ & $T_\text{ver}^y$ & $T_\text{ver}^z$ &
        $E_x$ & $E_y$ & $E_z$ & 
        $E_S$\\
        & & [a.u.] & [a.u.] & [a.u.] & & & &\\
        \midrule
        \ce{CH2O} & aug-cc-pVDZ & $450$ & $600$ & $650$ &$8 \cdot 10^{-6}$ & $8 \cdot 10^{-4}$ & $1 \cdot 10^{-3}$ & $1 \cdot 10^{-3}$\\
\ce{CO2}  & cc-pVDZ & $100$ & $100$ & $100$ &$8 \cdot 10^{-5}$ & $8 \cdot 10^{-5}$ & $3 \cdot 10^{-6}$ & $2 \cdot 10^{-4}$\\
 & aug-cc-pVDZ & $250$ & $250$ & $200$ &$6 \cdot 10^{-5}$ & $6 \cdot 10^{-5}$ & $3 \cdot 10^{-4}$ & $2 \cdot 10^{-3}$\\
  & aug-cc-pVTZ & $300$ & $300$ & $250$ &$2 \cdot 10^{-4}$ & $2 \cdot 10^{-4}$ & $1 \cdot 10^{-4}$ & $2 \cdot 10^{-3}$\\
\ce{H2O} & aug-cc-pVDZ & $150$ & $200$ & $300$ &$1 \cdot 10^{-5}$ & $2 \cdot 10^{-4}$ & $7 \cdot 10^{-5}$ & $3 \cdot 10^{-4}$\\
\ce{NH3} & aug-cc-pVDZ & $350$ & $300$ & $300$ &$4 \cdot 10^{-5}$ & $7 \cdot 10^{-4}$ & $7 \cdot 10^{-6}$ & $3 \cdot 10^{-3}$\\

        \bottomrule
    \end{tabular}
\end{table*}
\begin{table*}[ht]
    \centering
    \caption{Convergence times and corresponding errors of systems from RT-TDDFT calculations.}
    \label{tab:DFT}
    \begin{tabular}{lcccccccc}
        \toprule
         & basis &  
        $T_\text{ver}^x$ & $T_\text{ver}^y$ & $T_\text{ver}^z$ &
        $E_x$ & $E_y$ & $E_z$ & 
        $E_S$\\
        & & [a.u.] & [a.u.] & [a.u.] & & & &\\
        \midrule
        \ce{Be} & aug-ucc-pVTZ & $100$ & $100$ & $100$ &$1 \cdot 10^{-8}$ & $1 \cdot 10^{-8}$ & $1 \cdot 10^{-8}$ & $6 \cdot 10^{-6}$\\
\ce{C2H6} & aug-ucc-pVDZ & $750$ & $800$ & $550$ &$6 \cdot 10^{-4}$ & $7 \cdot 10^{-4}$ & $5 \cdot 10^{-4}$ & $8 \cdot 10^{-4}$\\
 & aug-ucc-pVTZ & $1000$ & $1000$ & $950$ &$5 \cdot 10^{-2}$ & $3 \cdot 10^{-2}$ & $2 \cdot 10^{-5}$ & $6 \cdot 10^{-3}$\\
\ce{C6H6} & aug-ucc-pVDZ & $1000$ & $1000$ & $550$ &$3 \cdot 10^{-2}$ & $4 \cdot 10^{-2}$ & $7 \cdot 10^{-4}$ & $4 \cdot 10^{-2}$\\
\ce{CH2O} & aug-ucc-pVDZ & $450$ & $650$ & $700$ &$4 \cdot 10^{-6}$ & $8 \cdot 10^{-4}$ & $1 \cdot 10^{-4}$ & $5 \cdot 10^{-4}$\\
 & aug-ucc-pVTZ & $650$ & $900$ & $1000$ &$2 \cdot 10^{-4}$ & $4 \cdot 10^{-4}$ & $2 \cdot 10^{-3}$ & $8 \cdot 10^{-4}$\\
\ce{CH3OH} & aug-ucc-pVDZ & $1000$ & $1000$ & $1000$ &$2 \cdot 10^{-1}$ & $8 \cdot 10^{-1}$ & $4 \cdot 10^{-2}$ & $8 \cdot 10^{-2}$\\
 & aug-ucc-pVTZ & $1000$ & $1000$ & $1000$ &$3 \cdot 10^{-1}$ & $8 \cdot 10^{-1}$ & $4 \cdot 10^{-1}$ & $2 \cdot 10^{-1}$\\
\ce{CH4} & aug-ucc-pVDZ & $200$ & $200$ & $200$ &$8 \cdot 10^{-4}$ & $7 \cdot 10^{-4}$ & $6 \cdot 10^{-4}$ & $2 \cdot 10^{-3}$\\
 & aug-ucc-pVTZ & $350$ & $350$ & $350$ &$4 \cdot 10^{-4}$ & $3 \cdot 10^{-4}$ & $3 \cdot 10^{-4}$ & $3 \cdot 10^{-3}$\\
\ce{CO2} & aug-ucc-pVDZ & $350$ & $350$ & $250$ &$4 \cdot 10^{-4}$ & $4 \cdot 10^{-4}$ & $1 \cdot 10^{-4}$ & $5 \cdot 10^{-4}$\\
\ce{H2O} & aug-ucc-pVDZ & $200$ & $250$ & $300$ &$2 \cdot 10^{-6}$ & $3 \cdot 10^{-6}$ & $3 \cdot 10^{-4}$ & $3 \cdot 10^{-4}$\\
\ce{H2} & aug-ucc-pVTZ & $100$ & $100$ & $100$ &$1 \cdot 10^{-7}$ & $1 \cdot 10^{-7}$ & $8 \cdot 10^{-8}$ & $5 \cdot 10^{-6}$\\
\ce{He} & aug-ucc-pVTZ & $100$ & $100$ & $100$ &$9 \cdot 10^{-8}$ & $9 \cdot 10^{-8}$ & $1 \cdot 10^{-7}$ & $9 \cdot 10^{-6}$\\
\ce{LiH} & aug-ucc-pVDZ & $100$ & $100$ & $300$ &$1 \cdot 10^{-5}$ & $3 \cdot 10^{-5}$ & $3 \cdot 10^{-4}$ & $3 \cdot 10^{-4}$\\
\ce{NH3} & aug-ucc-pVDZ & $350$ & $400$ & $300$ &$2 \cdot 10^{-4}$ & $4 \cdot 10^{-4}$ & $7 \cdot 10^{-4}$ & $1 \cdot 10^{-3}$\\

        \bottomrule
    \end{tabular}
\end{table*}

The fitting method reached the strict threshold for most systems, with a maximum spectral error of $E_S \leq 3\cdot 10^{-3}$. For all converged systems, the approximated functions for the dipole moment $\bar{\mu}_u(t)$ reliably reproduce its reference spectrum. The systems with very sparse spectra (\ce{He}, \ce{H2}, and \ce{Be}) converged instantly ($T_\text{ver} = 100\au$), providing approximated spectra indistinguishable from their reference spectra. In these cases, the fitting method achieved a speedup of 10 times compared to the max trajectory length of $T_\text{max} = 1000\au$, or
40 times compared to computing the reference spectra (using $4000\au$). The reduction in computational cost achieved by using the fitting method is relative to the desired spectral resolution. 
We would argue that the least unambiguous way to assess the speedup is to compare the convergence times $T^u_\text{ver}$ with the simulation time that would have been used if the fitting method was not used, i.e. the max trajectory length. 
Should the method converge at the preset max trajectory length, one may argue that no simulation time was spared. In this case, one still achieves arbitrary improvement in the spectral resolution.

Systems with relatively sparse spectra (\ce{CH4}, \ce{CO2}, \ce{H2O}, \ce{LiH}, and \ce{NH3}) also converged nicely with short dipole trajectories ($T_\text{ver}^u \leq 350\au$). As the spectral density increases, the fitting method struggles to converge. Systems like \ce{C2H6} and \ce{CH2O} only converged when using a double-zeta basis set, while the fitting of \ce{C6H6} and \ce{CH3OH} did not achieve errors below the low threshold.

\begin{figure}[ht]
    \centering
    \includegraphics[width=\linewidth]{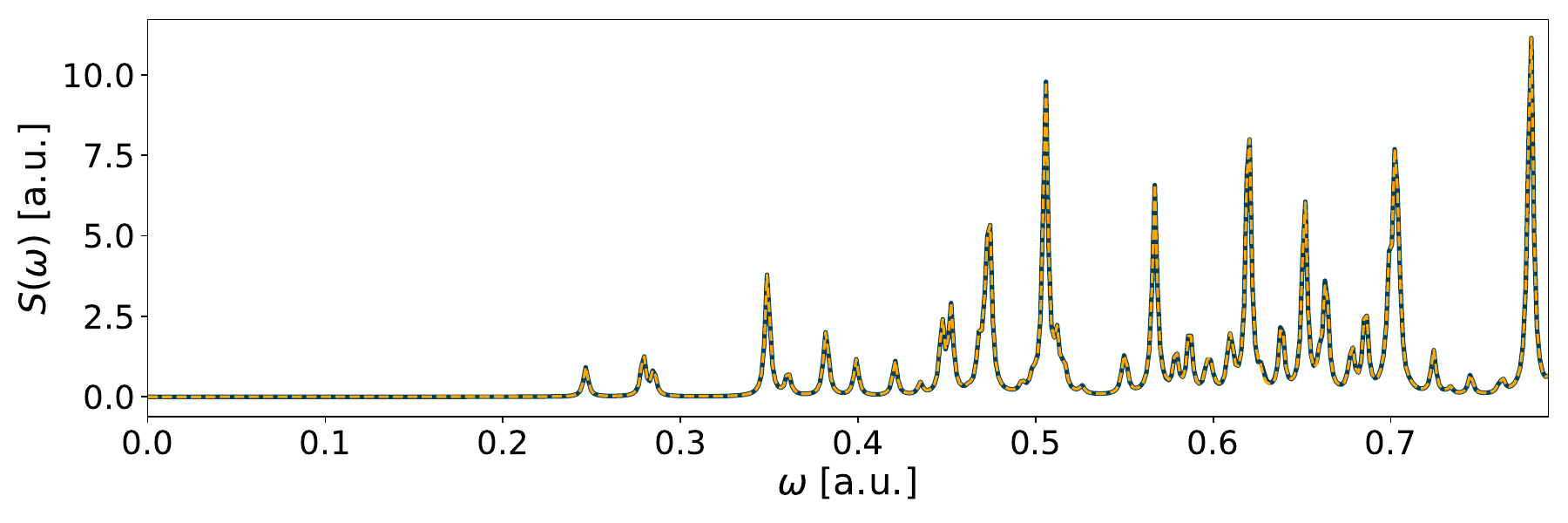}
    \caption{Spectrum of \ce{CH2O} using the aug-ucc-pVDZ basis in a RT-TDDFT simulation. The reference spectrum is in solid blue, while the yellow dashed line shows the spectrum of the fitted functions. The fitting error was $E_x = 4\cdot10^{-6}$, $E_y = 8\cdot10^{-4}$, and $E_z = 1\cdot10^{-4}$.}
    \label{fig:CH2O}
\end{figure}

The \ce{CH2O} molecule with a double-zeta basis set converged for both real-time methods. The spectra of the fit in both cases are nearly indistinguishable from their reference spectra. A comparison between the approximated and reference spectrum from RT-TDDFT calculations is shown in \cref{fig:CH2O}. The RT-TDDFT triple-zeta case nearly reached the error threshold ($E_z = 2\cdot 10^{-3}$), also providing a very low spectral error ($E_S = 8\cdot 10^{-4}$).

\begin{figure}[ht]
    \centering
    \includegraphics[width=\linewidth]{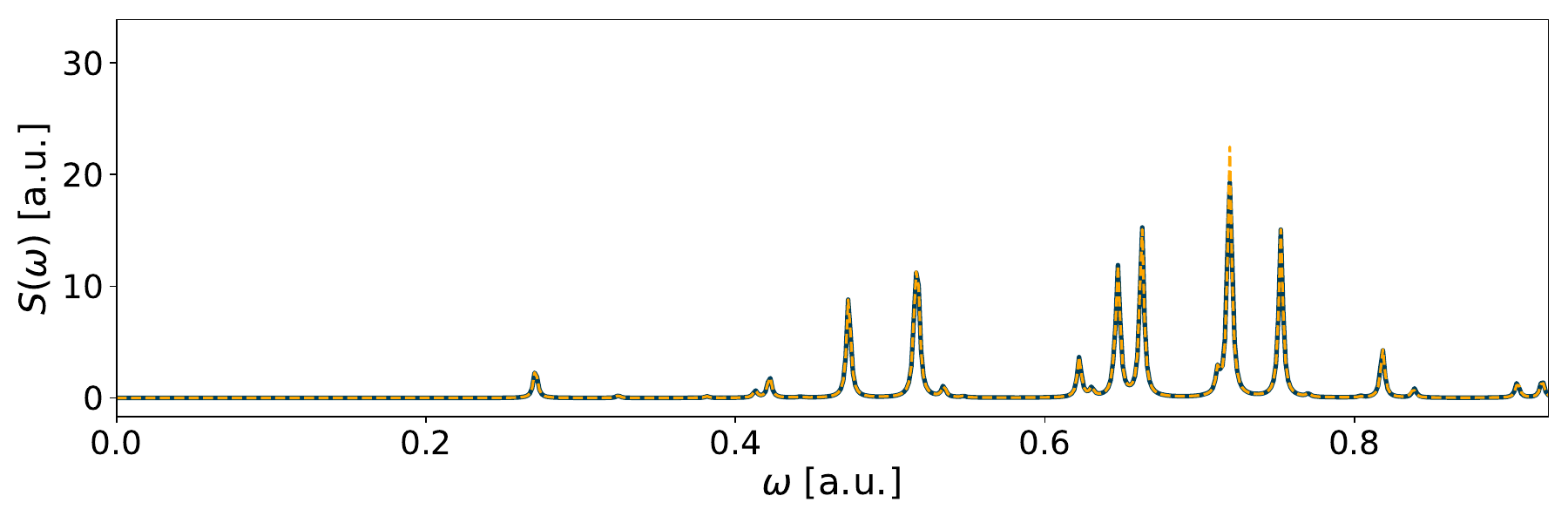}
    \caption{Spectrum of \ce{NH3} using the aug-cc-pVDZ basis in a RT-TDCIS simulation. The reference spectrum is in solid blue, while the yellow dashed line shows the spectrum of the converged fitted functions. This was the poorest approximated spectrum of all converged cases. The fitting error was $E_x = 4\cdot10^{-5}$, $E_y = 7\cdot10^{-4}$, and $E_z = 7\cdot10^{-6}$.}
    \label{fig:NH3}
\end{figure}

Among the converged systems, \ce{NH3} from RT-TDCIS calculations showed the largest error compared to its reference spectrum ($E_S = 3\cdot10^{-3}$). Its spectrum is shown in \cref{fig:NH3}, and was the approximated spectrum with the most visible deviation from its reference spectrum among the converged systems. 
The approximated spectrum shows a deviation in a peak at $\omega \approx 0.75\au$, but the rest of the peaks correspond well to the reference spectrum.

\begin{figure}[ht]
    \centering
    \includegraphics[width=\linewidth]{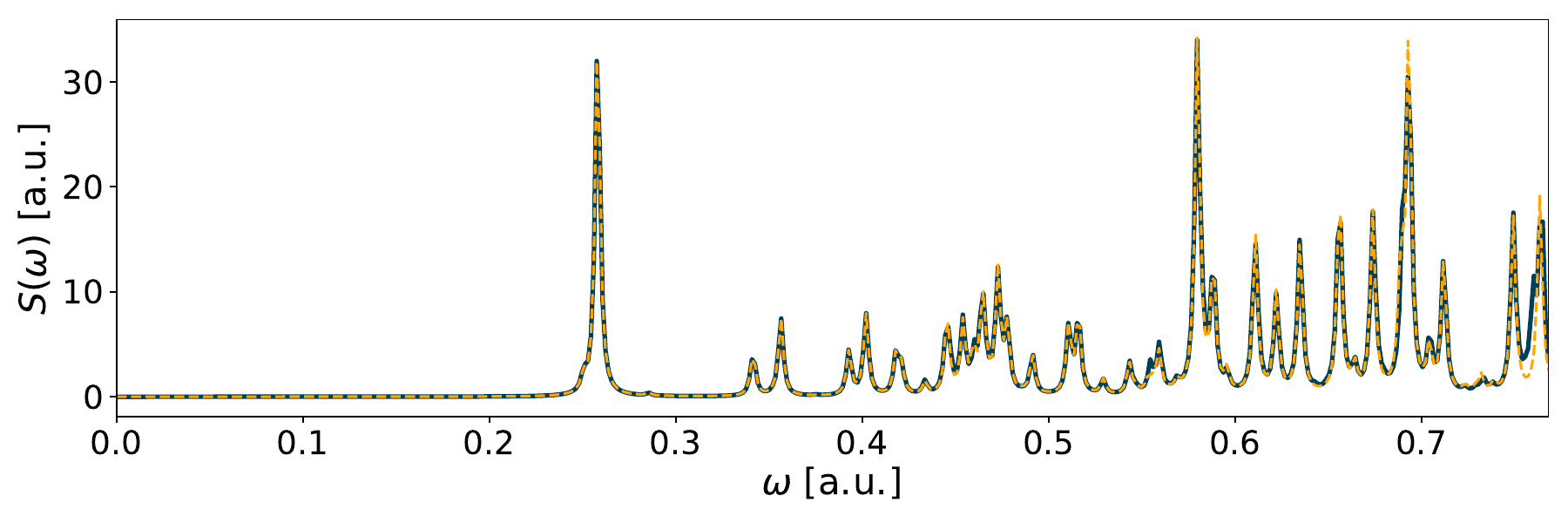}
    \caption{Spectrum of \ce{C6H6} using the aug-ucc-pVDZ basis in a RT-TDDFT simulation. The reference spectrum is in solid blue, while the yellow dashed line shows the spectrum of the fitted functions. The fitting error was $E_x = 3\cdot10^{-2}$, $E_y = 4\cdot10^{-2}$, and $E_z = 7\cdot10^{-4}$.}
    \label{fig:C6H6}
\end{figure}

The fitting method only partially converged for \ce{C6H6}, as well as \ce{C2H6} and \ce{CH2O} with triple-zeta basis, meaning that the error of the fit was below the set threshold in only one or two of the spatial directions. 
Still, the spectral error in all three cases is quite low. The result of the fitting of benzene is shown in \cref{fig:C6H6}, which had the largest spectral error ($E_S = 4\cdot 10^{-2}$) of the three. 

The trajectory length needed for the fitting method to converge strongly depends on the spectral density. We observed a trend in that the fitting becomes increasingly difficult as the spectral density increases. Increasing either the number of electrons in the system or the size of the basis set will in general require longer real-time simulations before the fitting method converges. The trend with increasing basis set size is clearly seen from the fitting of \ce{CO2} from RT-TDCIS calculations. The simulation using the cc-pVDZ basis set converges faster ($T_\text{ver}^u = 100\,\au$) than when using the larger basis sets like the aug-cc-pVDZ basis set ($T_\text{ver}^u \leq 250\,\au$) or aug-cc-pVTZ basis set ($T_\text{ver}^u \leq 300\,\au$).

\begin{figure}[ht]
    \centering
    \includegraphics[width=\linewidth]{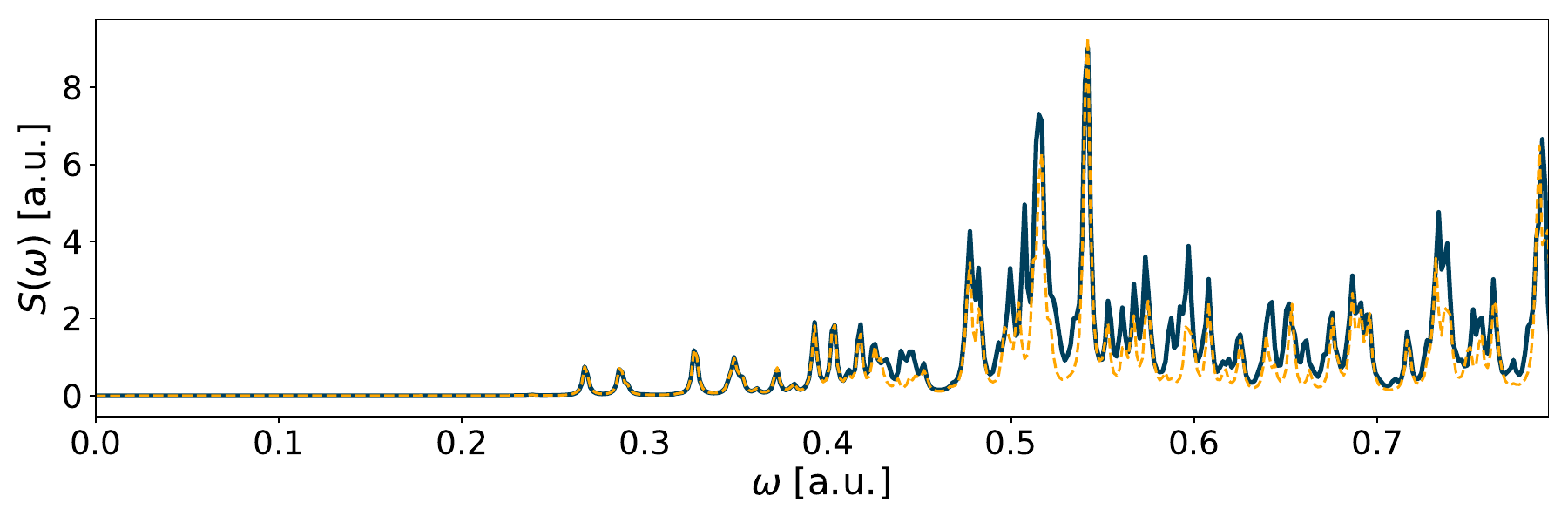}
    \caption{Spectrum of \ce{CH3OH} using the aug-ucc-pVTZ basis in a RT-TDDFT simulation. The reference spectrum is in solid blue, while the yellow dashed line shows the spectrum of the fitted functions. The fitting error was $E_x = 3\cdot10^{-1}$, $E_y = 8\cdot10^{-1}$, and $E_z = 4\cdot10^{-1}$.}
    \label{fig:CH3OH}
\end{figure}

Only the fit of \ce{CH3OH} did not get errors below the convergence threshold in any of the spatial directions. This was true for both the double and tripe-zeta basis (from RT-TDDFT calculations). The result using a triple-zeta basis set is shown in \cref{fig:CH3OH} and is the case with the highest error in the time domain, $E_u \sim 10^{-1}$. There is a significant deviation from the reference spectrum, though the main features are intact. Of all systems in this paper, this gave the worst approximation to the reference spectrum. Despite this, the spectrum $\bar{S}(\omega)$ still provides a reasonable coarse approximation.  

These results are promising in all cases as the converged fit seems to reproduce its reference spectrum reliably with only minor deviations in the peak intensities. The error of the fit $E_u$ also correlates with the spectral error, $E_S$. 
This predictability is crucial if the convergence criterion is used to automatically terminate real-time simulations. Our results also indicate that the convergence criterion used in this study is stricter than necessary. A slight relaxation in the criterion might lead to faster convergence without significantly impacting the quality of the approximated spectrum. 

For the estimated dipole moment, the frequencies and their corresponding linear coefficients are known. For a successful fit, one may therefore obtain the transition probability $\abs{\bra{0}\hat{\mu}_u\ket{n}}^2$ of a transition with energy $E_n - E_0$ directly from the linear coefficient, as $\abs{\bra{0}\hat{\mu}_u\ket{n}}^2 = B_n^u/(2\kappa)$. This could be used to calculate the oscillator strength and create stick spectra. However, estimated frequencies in different spatial directions but corresponding to the same transition will have a small error associated with the frequencies. In order to compute the oscillator strength, one would therefore have to assess which estimated frequencies across spatial directions correspond to the same transition.

The convergence of the dipole moment fitting depends primarily on the frequency estimation. When the fit does not converge, it follows that the Fourier-Padé approximant is not sufficiently converged to accurately capture the Bohr frequencies. The quality of the Fourier-Padé depends on the dipole trajectory length $t_N$ rather than the number of steps or step length~\cite{mattiat_efficient_2018}. However, there is no given final time $t_N$ ensuring convergence, the necessary trajectory length depends on the spectral density. High spectral density can cause the Fourier-Padé to fail, even for relatively long simulations. 
The general Padé approximant is prone to instabilities due to problems with near-degeneracy of the linear system. As pointed out by \citeauthor{cooper_short_2021}\cite{cooper_short_2021}, the Fourier-Padé used in real-time spectroscopy is known to struggle with dense spectra. 
The fitting method introduces a measure of the error $E_u$ which does not rely on any reference spectrum. This introduces a more reliable way of estimating the error in the approximated spectrum.

\subsection{Fitting using MO decomposition}
We assessed the performance of the fitting procedure used to extrapolate the components of the dipole moment of \ce{C6H6} decomposed to MO pairs $\bar{\mu}_u^{i a}$ in the RT-TDDFT calculation. Instead of creating a fitting function for each individual MO pair, which would increase the memory overhead, we clustered the components $\bar{\mu}_u^{i a}$ into groups of ten. These groups are formed so that the overall sparsity of spectra obtained for each cluster is maintained. This is accomplished by spreading the individual constituents of the cluster across the energy range. 
A fitted function of each cluster was then created from the sum over the MO pairs that the cluster contains. The total error is measured for the full dipole moment, $\mu_u (t)$. Some of the MO pairs were omitted entirely, making the MO decomposition work as a low-pass filter. When fitting components, the low-pass filter is therefore not needed. 

Fitting the decomposed signal, however, did not improve the convergence compared to when the full dipole moment was used for extrapolation. Simulations for both directions
$\mu_x$  and  $\mu_y$
reached the max trajectory length ($t = 1000\au$) without the fitting error going below the error threshold. The errors ($E_x = 1\cdot10^{-2}$ and $E_y = 1\cdot10^{-2}$) were only slightly lower compared to fitting without the MO decomposition. The last spatial direction $\mu_z$ converged at 
$T_\text{ver} = 650\au$ ($E_z = 4\cdot10^{-4}$),
which is somewhat slower than without MO decomposition. The spectral error was $E_S = 9\cdot 10^{-3}$, which corresponds to a low spectral error.

\begin{figure}[ht]
    \centering
    \includegraphics[width=\linewidth]{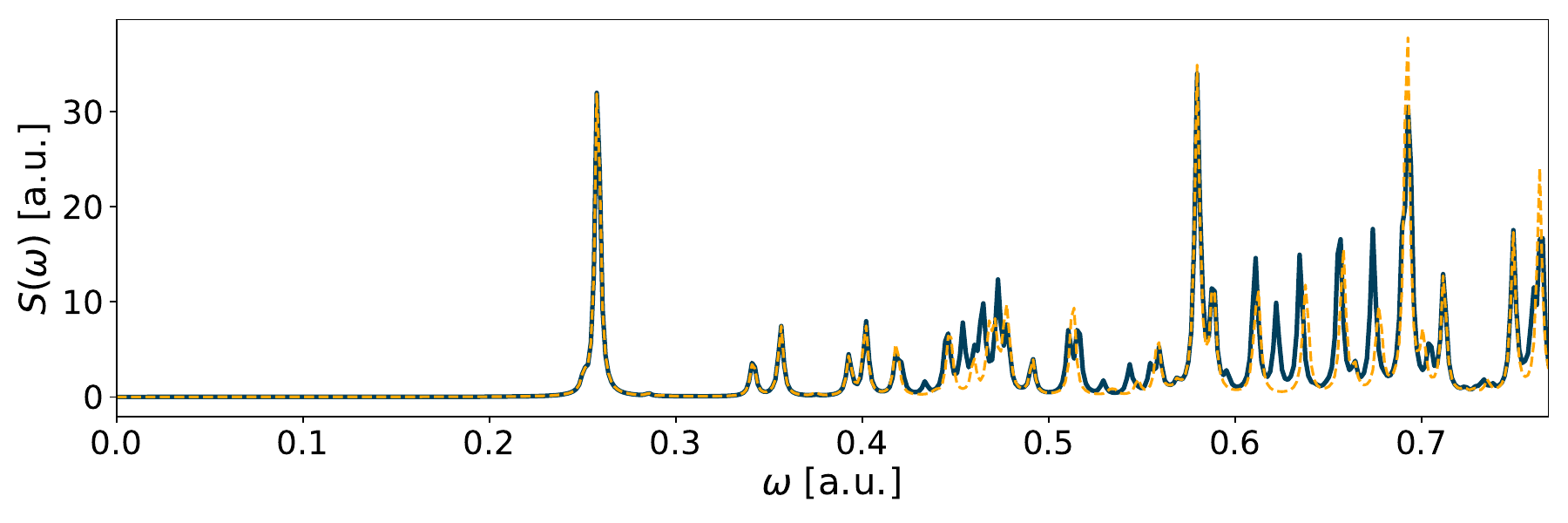}
    \caption{Spectrum of \ce{C6H6} using the aug-ucc-pVDZ basis in a RT-TDDFT simulation. The reference spectrum is in solid blue, while the yellow dashed line shows the spectrum of the fitted functions from 
    $T^u_\text{ver} = 600\au$.}
    \label{fig:full}
\end{figure}

\begin{figure}[ht]
    \centering
    \includegraphics[width=\linewidth]{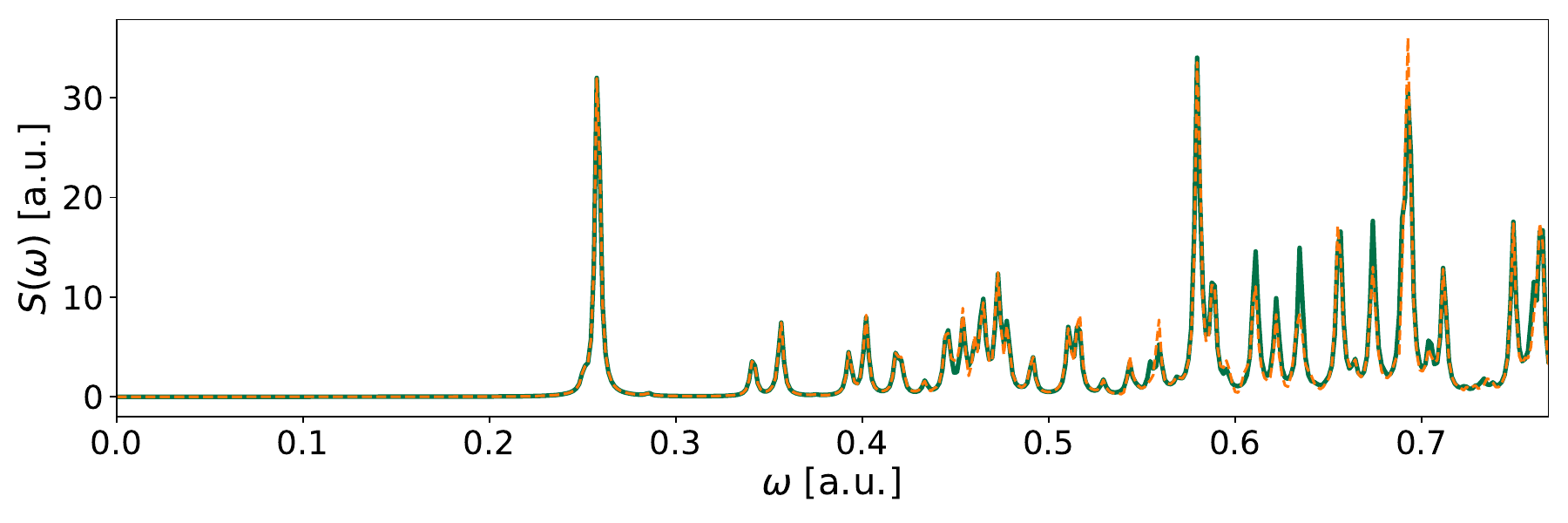}
    \caption{Spectrum of \ce{C6H6} using the aug-ucc-pVDZ basis in a RT-TDDFT simulation. The reference spectrum is in solid green, while the orange dashed line shows the spectrum of the fitted functions using molecular orbital decomposition from 
    $T^u_\text{ver} = 600\au$.}
    \label{fig:MO}
\end{figure}

Although the MO decomposition did not lead to accelerated convergence of the fitting method, we still observed improvements. For example the simulation with 
$T^u_\text{ver} = 600\au$
has a lower error of the fit for the decomposed dipole moment
($E_x = 6\cdot10^{-2}$, $E_y = 4\cdot10^{-2}$ and $E_z = 1\cdot10^{-3}$)
for all spatial directions compared to the fit on the full dipole moment, 
($E_x = E_y = 3\cdot10^{-1}$ and $E_z = 2\cdot10^{-3}$).
The decomposed fit in \cref{fig:MO} ($E_S = 3\cdot10^{-2}$) is visibly improved compared to the fit using the full dipole moment in \cref{fig:full} ($E_S = 2\cdot10^{-1}$). 


It is important to note that the scope of our testing of the fitting method using MO decomposition was limited. Previous success using the Fourier-Padé approximant in combination with the MO decomposition on RT-TDDFT data suggests that this in many cases is very effective\cite{bruner_accelerated_2016}.
Our study, however, raises cause for caution regarding the use of the Fourier-Padé approximant:
The Padé \textit{can} struggle, even when using MO decomposition. The unknown amount of error introduced to the final spectrum by this procedure remains an open problem that the user should be aware of when analyzing spectra with the Fourier-Padé method with the MO decomposition.

The particular form of the components  $\mu_u^{p q}$ is also very dependent on the quantum mechanical method used to compute the time-dependent wave function. Using MO decomposition on the electric dipole moment from RT-TDCCSD calculations leads to large overlaps in frequencies among different components\cite{hauge_extrapolating_2021}. The usefulness of such decomposition might vary between the different quantum mechanical frameworks.

\section{Conclusion}
We have developed a novel method for creating functions approximating the electric dipole moment from real-time calculations. The fitted functions for the dipole moment in the three spatial directions can then be used to produce absorption spectra with arbitrary high resolution. Real-time calculations of absorption spectra require the use of the discrete Fourier transforms, demanding long simulation times to obtain high spectral resolution. In our work, we have shown that the length of the real-time simulations, and hence the computational cost, can be greatly reduced by the developed fitting method. 

We introduced a quantitative error measure to evaluate the convergence of the fit. For all systems in this work, a converged fit reliably reproduced the reference spectrum from long real-time calculations. 
A convergence criterion of $10^{-3}$ seems to be quite strict, and further studies should be conducted to investigate the impact of slightly higher errors on the estimated spectrum.
In order to reduce the computational cost of calculating absorption spectra, the real-time calculations should be automatically terminated once the convergence criterion is reached.

In this work, we set the verification window to be $25\%$ of the available dipole trajectory. The critical step of the method is determining the frequencies, which always uses all available data. For the linear optimization, the verification window should include an entire period of the smallest frequency in the signal, as an insufficiently large verification window may lead to misleading error estimates. In future work, the verification window should depend on an estimate of the smallest frequency in the signal based on differences in the molecular orbital energies.

The fitting method converged with as little as $100\au$ long trajectories in time for systems with sparse spectra. Convergence slows down as spectral density increases, even leading to failure of convergence in some cases. The current version of the fitting method shows encouraging results for smaller systems, although aspects of the method require further investigation. 

Our testing of the fitting method using molecular orbital decomposition of a single system gave mixed results. The decomposition did not enable the fit to meet the convergence criterion, although we observed improvements in the approximated spectrum. This motivates the need for further investigations.

An apparent weakness of the current implementation is the way of estimating frequencies. Future versions should not rely on the Fourier-Padé but rather investigate other methods of estimating frequencies. This could include other methods for harmonic inversion or letting the function form of the fitted dipole moment be a truncated Fourier series based on an estimation of the fundamental frequency. 
The same frequencies can appear in all spatial directions, which could be exploited to improve the frequency estimation. In particular, in cases where the frequencies are successfully estimated in one spatial direction, knowledge of these existing frequencies could be used to alleviate the search in the other spatial directions with potentially higher spectral density. 
Improving the frequency estimation is crucial for stabilizing the fitting method for systems with high spectral density. 

This work has focused on the Dirac delta impulse, though the general fitting algorithm may be used on systems with any type of external field. Using a laser pulse targeting a specific spectral region may provide both an upper and lower bound when estimating the frequencies. Any \emph{a priori} information about the frequencies should be exploited by the fitting algorithm.
Additionally, the Dirac delta impulse targets \emph{all} excitation energies, maximizing the spectral density.
It is not unlikely that a narrow-band laser pulse would somewhat alleviate the fitting process by reducing the spectral density.  

\begin{acknowledgement}
    This work was supported by the Research Council of Norway through its Centres of Excellence scheme, project number 262695, and its Mobility Grant scheme (project nos. 301864 and 314814).
    The simulations were performed on resources provided by Sigma2---the National Infrastructure for High Performance Computing
    and Data Storage in Norway, Grant No.~NN4654K. T.~B.~P.~acknowledges the support of the Centre for Advanced Study in Oslo,
    Norway, which funded and hosted the CAS research project \emph{Attosecond Quantum Dynamics Beyond the Born-Oppenheimer Approximation}
    during the academic year 2021-2022. In addition, this project received funding from the European Union’s Horizon 2020 research and innovation program under the Marie Skłodowska-Curie Grant Agreement No. 945478 (SASPRO2), and the Slovak Research and Development Agency (Grant No. APVV-21-0497).
\end{acknowledgement}

\begin{suppinfo}
    Molecular geometries and HOMO energies for all systems. Comparisons between the spectra of filtered versus unfiltered dipole moments. Spectra of the fitted functions of the dipole moments compared with the corresponding reference spectrum. 
\end{suppinfo}
\bibliography{main}

\end{document}


\section{Molecular geometries}

The molecular geometries of all systems used in the numerical study are listed in \cref{tab:C2H6,tab:C6H6,tab:CH2O,tab:CH3OH,tab:CH4,tab:CO2,tab:H2O,tab:H2,tab:LiH,tab:NH3}. The same geometries were used in both the real-time time-dependent density-functional theory (RT-TDDFT) and the real-time time-dependent configuration interaction singles (RT-TDCIS) calculations.

\begin{table}
\centering
\caption{Molecular geometry of \ce{C2H6}.}
\label{tab:C2H6}
\begin{tabular}{lrrr}
	\toprule
	& $x$ [\AA] & $y$ [\AA] & $z$ [\AA]\\
	\midrule
	C & $0.0000$ & $0.0000$ & $0.7610$\\
	 & $0.0000$ & $0.0000$ & $-0.7610$\\
	H & $0.0000$ & $1.0233$ & $1.1616$\\
	 & $-0.8862$ & $-0.5116$ & $1.1616$\\
	 & $0.8862$ & $-0.5116$ & $1.1616$\\
	 & $0.0000$ & $-1.0233$ & $-1.1616$\\
	 & $-0.8862$ & $0.5116$ & $-1.1616$\\
	 & $0.8862$ & $0.5116$ & $-1.1616$\\
	\bottomrule
\end{tabular}
\end{table}
\begin{table}
\centering
\caption{Molecular geometry of \ce{C6H6}.}
\label{tab:C6H6}
\begin{tabular}{lrrr}
	\toprule
    & $x$ [\AA] & $y$ [\AA] & $z$ [\AA]\\
	\midrule
	C & $0.0000$ & $1.3949$ & $0.0000$\\
	 & $1.2080$ & $0.6975$ & $0.0000$\\
	 & $1.2080$ & $-0.6975$ & $0.0000$\\
	 & $0.0000$ & $-1.3949$ & $0.0000$\\
	 & $-1.2080$ & $-0.6975$ & $0.0000$\\
	 & $-1.2080$ & $0.6975$ & $0.0000$\\
	H & $0.0000$ & $2.4858$ & $0.0000$\\
	 & $2.1528$ & $1.2429$ & $0.0000$\\
	 & $2.1528$ & $-1.2429$ & $0.0000$\\
	 & $0.0000$ & $-2.4858$ & $0.0000$\\
	 & $-2.1528$ & $-1.2429$ & $0.0000$\\
	 & $-2.1528$ & $1.2429$ & $0.0000$\\
	\bottomrule
\end{tabular}
\end{table}
\begin{table}
\centering
\caption{Molecular geometry of \ce{CH2O}.}
\label{tab:CH2O}
\begin{tabular}{lrrr}
\toprule
& $x$ [\AA] & $y$ [\AA] & $z$ [\AA]\\
\midrule
    O &     $0.0000$ &      $ 0.0000$ &      $ 0.6751$ \\
    C &     $0.0000$ &      $ 0.0000$ &      $-0.5280$ \\
    H &     $0.0000$ &      $ 0.9462$ &      $-1.1162$ \\
     &     $0.0000$ &      $-0.9462$ &      $-1.1162$ \\
\bottomrule
\end{tabular}
\end{table}
\begin{table}
\centering
\caption{Molecular geometry of \ce{CH3OH}.}
\label{tab:CH3OH}
\begin{tabular}{lrrr}
	\toprule
	& $x$ [\AA] & $y$ [\AA] & $z$ [\AA]\\
	\midrule
    C &    $-0.0463$ &      $ 0.6610$ &      $ 0.0000$ \\
    O &    $-0.0463$ &      $-0.7546$ &      $ 0.0000$ \\
    H &    $-1.0964$ &      $ 0.9762$ &      $ 0.0000$ \\
     &    $ 0.4402$ &      $ 1.0772$ &      $ 0.8975$ \\
     &    $ 0.4402$ &      $ 1.0772$ &      $-0.8975$ \\
     &    $ 0.8645$ &      $-1.0602$ &      $ 0.0000$ \\
\bottomrule
\end{tabular}
\end{table}
\begin{table}
\centering
\caption{Molecular geometry of \ce{CH4}.}
\label{tab:CH4}
\begin{tabular}{lrrr}
	\toprule
	& $x$ [\AA] & $y$ [\AA] & $z$ [\AA]\\
	\midrule
    C &    $ 0.0000$ &      $ 0.0000$ &      $ 0.0000$ \\
    H &    $ 0.6328$ &      $ 0.6328$ &      $ 0.6328$ \\
     &    $-0.6328$ &      $-0.6328$ &      $ 0.6328$ \\
     &    $-0.6328$ &      $ 0.6328$ &      $-0.6328$ \\
     &    $ 0.6328$ &      $-0.6328$ &      $-0.6328$ \\
\bottomrule
\end{tabular}
\end{table}

Each system's highest occupied molecular orbital energy is listed in \cref{tab:HOMO_DFT,tab:HOMO_CIS}, for the calculations using RT-TDDFT and RT-TDCIS respectively.

\begin{table}
\centering
\caption{Molecular geometry of \ce{CO2}.}
\label{tab:CO2}
\begin{tabular}{lrrr}
	\toprule
	& $x$ [\AA] & $y$ [\AA] & $z$ [\AA]\\
	\midrule
    C &     $0.0000$ &       $0.0000$ &      $ 0.0000$ \\
    O &     $0.0000$ &       $0.0000$ &      $ 1.1363$ \\
     &     $0.0000$ &       $0.0000$ &      $-1.1363$ \\
\bottomrule
\end{tabular}
\end{table}
\begin{table}
\centering
\caption{Molecular geometry of \ce{H2O}.}
\label{tab:H2O}
\begin{tabular}{lrrr}
	\toprule
	& $x$ [\AA] & $y$ [\AA] & $z$ [\AA]\\
	\midrule
    O &    $ 0.0000$ &      $ 0.0000$ &      $-0.0656$ \\
    H &    $ 0.0000$ &      $ 0.7567$ &      $ 0.5203$ \\
     &    $ 0.0000$ &      $-0.7567$ &      $ 0.5203$ \\
\bottomrule
\end{tabular}
\end{table}
\begin{table}
\centering
\caption{Molecular geometry of \ce{LiH}.}
\label{tab:LiH}
\begin{tabular}{lrrr}
	\toprule
    & $x$ [\AA] & $y$ [\AA] & $z$ [\AA]\\
    \midrule
   Li &    $ 0.0000$ &       $0.0000$ &       $0.0000$ \\
   H  &    $ 0.0000$ &       $0.0000$ &       $1.6299$ \\
\bottomrule
\end{tabular}
\end{table}
\begin{table}
\centering
\caption{Molecular geometry of \ce{H2}.}
\label{tab:H2}
\begin{tabular}{lrrr}
	\toprule
    & $x$ [\AA] & $y$ [\AA] & $z$ [\AA]\\
    \midrule
    H &    $ 0.0000$ &       $0.0000$ &       $0.0000$ \\
     &    $ 0.0000$ &       $0.0000$ &       $0.7408$ \\
\bottomrule
\end{tabular}
\end{table}
\begin{table}
\centering
\caption{Molecular geometry of \ce{NH3}.}
\label{tab:NH3}
\begin{tabular}{lrrr}
	\toprule
    & $x$ [\AA] & $y$ [\AA] & $z$ [\AA]\\
    \midrule
    N &    $ 0.0000$ &      $ 0.0000$ &      $ 0.1064$ \\
    H &    $ 0.0000$ &      $ 0.9335$ &      $-0.2482$ \\
     &    $ 0.8084$ &      $-0.4667$ &      $-0.2482$ \\
     &    $-0.8084$ &      $-0.4667$ &      $-0.2482$ \\
\bottomrule
\end{tabular}
\end{table}

\begin{table*}
    \centering
    \caption{Highest occupied molecular orbital energy of the systems using DFT calculations.}
    \label{tab:HOMO_DFT}
    \begin{tabular}{ccccccccc}
        \toprule
        & basis & $\epsilon_\text{HOMO}$ [a.u.]\\
        \midrule
        \ce{Be} & aug-ucc-pVTZ & $-0.2386$\\
        \ce{C2H6} & aug-ucc-pVDZ & $-0.3506$\\
          & aug-ucc-pVTZ & $-0.3509$\\
        \ce{C6H6} & aug-ucc-pVDZ & $-0.2681$\\
        \ce{CH2O} & aug-ucc-pVDZ & $-0.2887$\\
          & aug-ucc-pVTZ & $-0.2894$\\
        \ce{CH3OH} & aug-ucc-pVDZ & $-0.2925$\\
          & aug-ucc-pVTZ & $-0.2935$\\
        \ce{CH4} & aug-ucc-pVDZ & $-0.4024$\\
          & aug-ucc-pVTZ & $-0.4026$\\
        \ce{CO2} & aug-ucc-pVDZ & $-0.3947$\\
        \ce{H2O} & aug-ucc-pVDZ & $-0.3333$\\
        \ce{H2} & aug-ucc-pVTZ & $-0.4412$\\
        \ce{He} & aug-ucc-pVTZ & $-0.6695$\\
        \ce{LiH} & aug-ucc-pVDZ & $-0.1992$\\
        \ce{NH3} & aug-ucc-pVDZ & $-0.2807$\\
        \bottomrule
    \end{tabular}
\end{table*}
\begin{table*}
    \centering
    \caption{Highest occupied molecular orbital energy of the systems using CIS calculations.}
    \label{tab:HOMO_CIS}
    \begin{tabular}{ccccccccc}
        \toprule
        & basis & $\epsilon_\text{HOMO}$ [a.u.]\\
        \midrule
        \ce{CH2O} & aug-cc-pVDZ & $-0.4345$\\
        \ce{CO2} & cc-pVDZ & $-0.5372$\\
          & aug-cc-pVDZ & $-0.5459$\\
          & aug-cc-pVTZ & $-0.5463$\\
        \ce{H2O} & aug-cc-pVDZ & $-0.5095$\\
        \ce{NH3} & aug-cc-pVDZ & $-0.4253$\\
        \bottomrule
    \end{tabular}
\end{table*}

\section{Filtering the dipole moment}
The reference spectra are filtered using a low-pass filter. The comparisons between the filtered and unfiltered spectra are shown in
\cref{fig:CH2O_CIS_aug-cc-pVDZ_filter,fig:CO2_CIS_aug-cc-pVDZ_filter,fig:CO2_CIS_cc-pVDZ_filter,fig:CO2_CIS_aug-cc-pVTZ_filter,fig:H2O_CIS_aug-cc-pVDZ_filter,fig:NH3_CIS_aug-cc-pVDZ_filter,fig:Be_DFT_aug-ucc-pVTZ_filter,fig:C2H6_DFT_aug-ucc-pVDZ_filter,fig:C2H6_DFT_aug-ucc-pVTZ_filter,fig:CH2O_DFT_aug-ucc-pVDZ_filter,fig:CH2O_DFT_aug-ucc-pVTZ_filter,fig:CH3OH_DFT_aug-ucc-pVDZ_filter,fig:CH3OH_DFT_aug-ucc-pVTZ_filter,fig:CH4_DFT_aug-ucc-pVDZ_filter,fig:CH4_DFT_aug-ucc-pVTZ_filter,fig:C6H6_DFT_aug-ucc-pVDZ_filter,fig:CO2_DFT_aug-ucc-pVDZ_filter,fig:H2O_DFT_aug-ucc-pVDZ_filter,fig:H2_DFT_aug-ucc-pVTZ_filter,fig:He_DFT_aug-ucc-pVTZ_filter,fig:LiH_DFT_aug-ucc-pVDZ_filter,fig:NH3_DFT_aug-ucc-pVDZ_filter}, contrasting the filtered spectra (dark blue line) with the parts which have been filtered (faded blue line). The cut-off frequency was set to $\omega_\text{max} = 4\au$, which is safely higher than what could be expected values for ionization energies.  As can be seen from the figures, the low-pass filter does not provide a clean cut in the spectrum. Peaks around $\omega_\text{max}$ have lower intensity, also on the left side in the spectra. 

The figures showing the filtering of spectra from the RT-TDDFT calculations are shown in
\cref{fig:Be_DFT_aug-ucc-pVTZ_filter,fig:C2H6_DFT_aug-ucc-pVDZ_filter,fig:C2H6_DFT_aug-ucc-pVTZ_filter,fig:CH2O_DFT_aug-ucc-pVDZ_filter,fig:CH2O_DFT_aug-ucc-pVTZ_filter,fig:CH3OH_DFT_aug-ucc-pVDZ_filter,fig:CH3OH_DFT_aug-ucc-pVTZ_filter,fig:CH4_DFT_aug-ucc-pVDZ_filter,fig:CH4_DFT_aug-ucc-pVTZ_filter,fig:C6H6_DFT_aug-ucc-pVDZ_filter,fig:CO2_DFT_aug-ucc-pVDZ_filter,fig:H2O_DFT_aug-ucc-pVDZ_filter,fig:H2_DFT_aug-ucc-pVTZ_filter,fig:He_DFT_aug-ucc-pVTZ_filter,fig:LiH_DFT_aug-ucc-pVDZ_filter,fig:NH3_DFT_aug-ucc-pVDZ_filter}, 
while 
\cref{fig:CH2O_CIS_aug-cc-pVDZ_filter,fig:CO2_CIS_aug-cc-pVTZ_filter,fig:CO2_CIS_aug-cc-pVDZ_filter,fig:CO2_CIS_cc-pVDZ_filter,fig:H2O_CIS_aug-cc-pVDZ_filter,fig:NH3_CIS_aug-cc-pVDZ_filter}
shows the filtering of spectra from the RT-TDCIS calculations. 

\begin{figure*}
    \includegraphics[width=\linewidth]
    {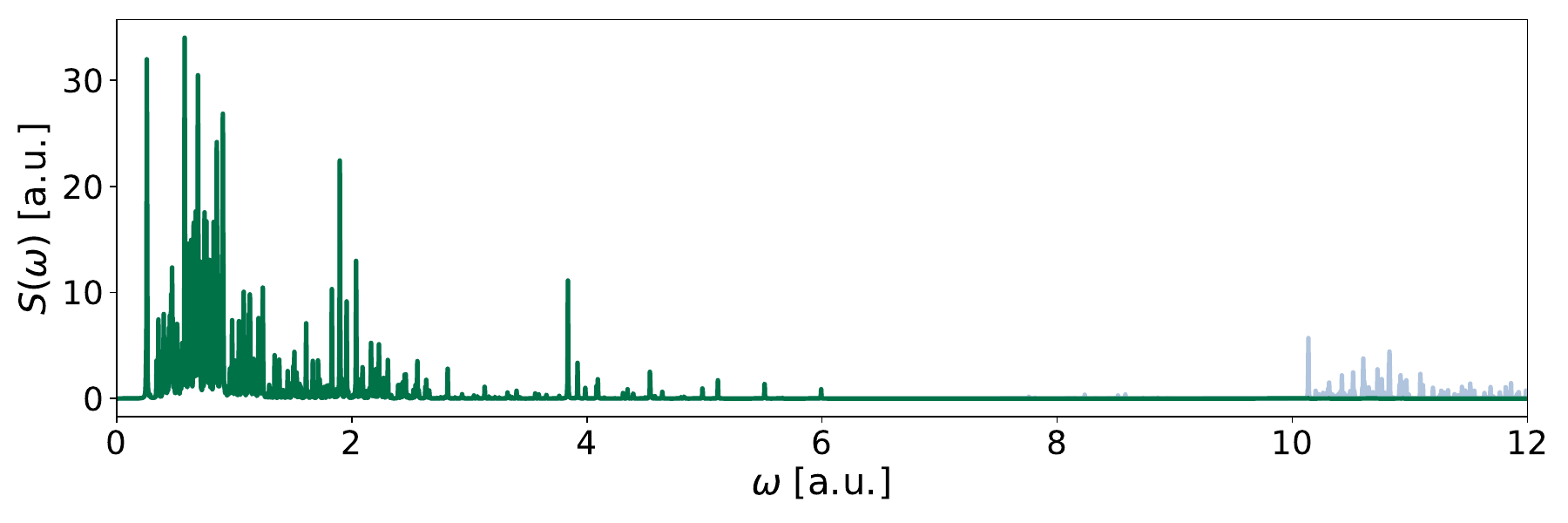}
    \caption{Effects of discarding high-energy molecular orbital transitions 
    on the absorption spectrum of \ce{C6H6} using the aug-ucc-pVDZ basis in a RT-TDDFT simulation. The dark green line shows the spectrum of the sum of the molecular orbital pair components, while the faded line shows the full unfiltered spectrum.}
    \label{fig:C6H6_dft_aug-ucc-pVDZ_MO_filter}
\end{figure*}

The effects of the discarding high-energy molecular orbital transitions on the spectrum of \ce{C6H6} using the aug-ucc-pVDZ basis set in a RT-TDDFT calculation is shown in 
\cref{fig:C6H6_dft_aug-ucc-pVDZ_MO_filter}. Here, the spectrum of the molecular orbital components (dark green line) is compared to the full unfiltered spectrum (faded blue line). 

\filtered{Be}{DFT}{aug-ucc-pVTZ}
\filtered{C2H6}{DFT}{aug-ucc-pVDZ}
\filtered{C2H6}{DFT}{aug-ucc-pVTZ}
\filtered{C6H6}{DFT}{aug-ucc-pVDZ}
\filtered{CH2O}{DFT}{aug-ucc-pVDZ}
\filtered{CH2O}{DFT}{aug-ucc-pVTZ}
\filtered{CH3OH}{DFT}{aug-ucc-pVDZ}
\filtered{CH3OH}{DFT}{aug-ucc-pVTZ}
\filtered{CH4}{DFT}{aug-ucc-pVDZ}
\filtered{CH4}{DFT}{aug-ucc-pVTZ}
\filtered{CO2}{DFT}{aug-ucc-pVDZ}
\filtered{H2O}{DFT}{aug-ucc-pVDZ}
\filtered{H2}{DFT}{aug-ucc-pVTZ}
\filtered{He}{DFT}{aug-ucc-pVTZ}
\filtered{LiH}{DFT}{aug-ucc-pVDZ}
\filtered{NH3}{DFT}{aug-ucc-pVDZ}

\filtered{CH2O}{CIS}{aug-cc-pVDZ}
\filtered{CO2}{CIS}{cc-pVDZ}
\filtered{CO2}{CIS}{aug-cc-pVDZ}
\filtered{CO2}{CIS}{aug-cc-pVTZ}
\filtered{H2O}{CIS}{aug-cc-pVDZ}
\filtered{NH3}{CIS}{aug-cc-pVDZ}

\section{Spectra of the approximated functions}

The figures in 
\cref{fig:CH2O_CIS_aug-cc-pVDZ,fig:CO2_CIS_aug-cc-pVTZ,fig:CO2_CIS_aug-cc-pVDZ,fig:CO2_CIS_cc-pVDZ,fig:H2O_CIS_aug-cc-pVDZ,fig:NH3_CIS_aug-cc-pVDZ}
compares the spectrum of the fitted function (yellow dashed line) with its corresponding reference spectrum (solid dark blue line). The convergence times $T_\text{ver}^u$ for $u = x, y, z$ are listed in the figures. The maximum trajectory length was set to $T_\text{ver}^u = 1000\au$, such that spectra with $T_\text{ver}^u = 1000\au$ in any direction are deemed not converged spectra.

The results of the fitting of the time-dependent dipole moment from RT-TDDFT calculations are shown in 
\cref{fig:Be_DFT_aug-ucc-pVTZ,fig:C2H6_DFT_aug-ucc-pVDZ,fig:C2H6_DFT_aug-ucc-pVTZ,fig:CH2O_DFT_aug-ucc-pVDZ,fig:CH2O_DFT_aug-ucc-pVTZ,fig:CH3OH_DFT_aug-ucc-pVDZ,fig:CH3OH_DFT_aug-ucc-pVTZ,fig:CH4_DFT_aug-ucc-pVDZ,fig:CH4_DFT_aug-ucc-pVTZ,fig:C6H6_DFT_aug-ucc-pVDZ,fig:CO2_DFT_aug-ucc-pVDZ,fig:H2O_DFT_aug-ucc-pVDZ,fig:H2_DFT_aug-ucc-pVTZ,fig:He_DFT_aug-ucc-pVTZ,fig:LiH_DFT_aug-ucc-pVDZ,fig:NH3_DFT_aug-ucc-pVDZ}, 
while 
\cref{fig:CH2O_CIS_aug-cc-pVDZ,fig:CO2_CIS_aug-cc-pVTZ,fig:CO2_CIS_aug-cc-pVDZ,fig:CO2_CIS_cc-pVDZ,fig:H2O_CIS_aug-cc-pVDZ,fig:NH3_CIS_aug-cc-pVDZ}
shows the resulting spectra from fitting the dipole moment from RT-TDCIS calculations.

\begin{figure*}
    \centering
    \includegraphics[width=\linewidth]
    {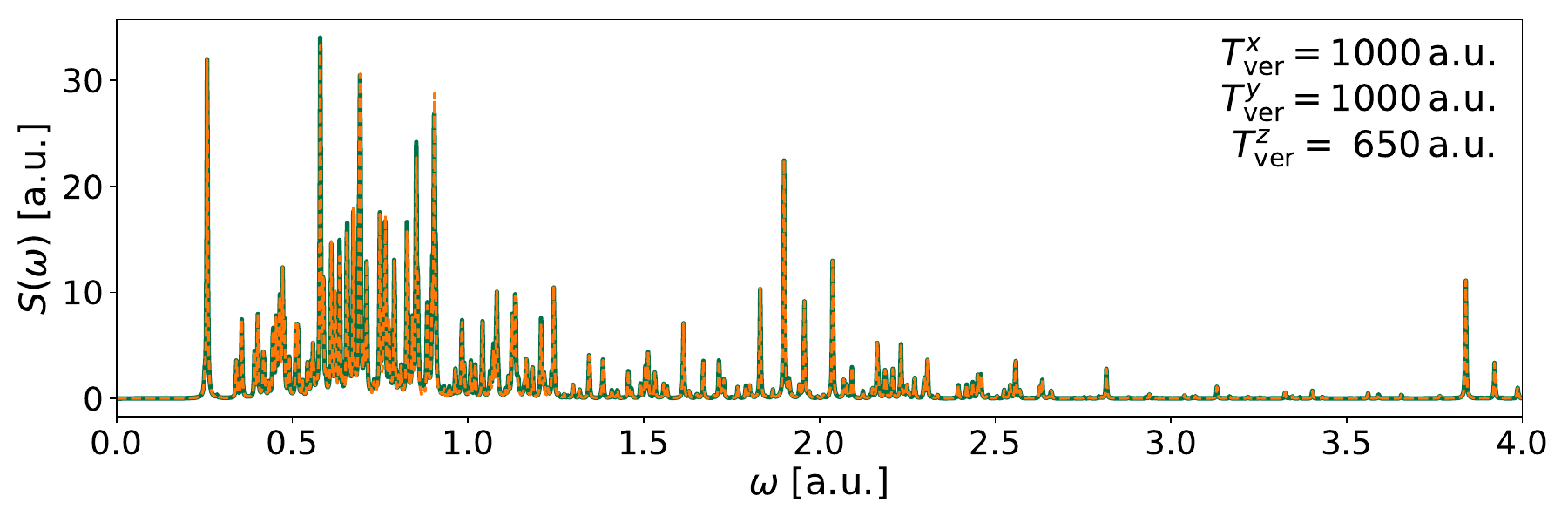}
    \caption{Spectrum of \ce{C6H6} using the aug-ucc-pVDZ basis in a RT-TDDFT simulation, discarding high-energy molecular orbital components. The spectrum of the approximated dipole moment using molecular orbital decomposition is shown by the dashed orange line, while the solid dark green is the reference spectrum.}
    \label{fig:C6H6_dft_aug-ucc-pVDZ_MO}
\end{figure*}

The resulting spectrum from fitting the electric dipole moment using molecular orbital decomposition (dashed orange line) is compared to its reference spectrum (solid dark green line) in \cref{fig:C6H6_dft_aug-ucc-pVDZ_MO}. This case uses \ce{C6H6} with the aug-ucc-pVDZ basis set, from RT-TDDFT calculations.

\spectrum{Be}{DFT}{aug-ucc-pVTZ}
\spectrum{C2H6}{DFT}{aug-ucc-pVDZ}
\spectrum{C2H6}{DFT}{aug-ucc-pVTZ}
\spectrum{C6H6}{DFT}{aug-ucc-pVDZ}
\spectrum{CH2O}{DFT}{aug-ucc-pVDZ}
\spectrum{CH2O}{DFT}{aug-ucc-pVTZ}
\spectrum{CH3OH}{DFT}{aug-ucc-pVDZ}
\spectrum{CH3OH}{DFT}{aug-ucc-pVTZ}
\spectrum{CH4}{DFT}{aug-ucc-pVDZ}
\spectrum{CH4}{DFT}{aug-ucc-pVTZ}
\spectrum{CO2}{DFT}{aug-ucc-pVDZ}
\spectrum{H2O}{DFT}{aug-ucc-pVDZ}
\spectrum{H2}{DFT}{aug-ucc-pVTZ}
\spectrum{He}{DFT}{aug-ucc-pVTZ}
\spectrum{LiH}{DFT}{aug-ucc-pVDZ}
\spectrum{NH3}{DFT}{aug-ucc-pVDZ}

\spectrum{CH2O}{CIS}{aug-cc-pVDZ}
\spectrum{CO2}{CIS}{cc-pVDZ}
\spectrum{CO2}{CIS}{aug-cc-pVDZ}
\spectrum{CO2}{CIS}{aug-cc-pVTZ}
\spectrum{H2O}{CIS}{aug-cc-pVDZ}
\spectrum{NH3}{CIS}{aug-cc-pVDZ}